\documentclass[twocolumn,superscriptaddress,longbibliography,10pt,aps,pra,noshowpacs]{revtex4-2}
\usepackage{amsmath}
\usepackage{amsfonts}
\usepackage{amssymb}
\usepackage{ulem}
\usepackage[dvipsnames]{xcolor}
\usepackage{graphicx}
\usepackage{hyperref}
\hypersetup{
	colorlinks=true,
	citecolor=blue,
	linkcolor=blue,
	urlcolor=blue}
\newcommand{\p}{\partial}
\newcommand{\mb}{\mathbf}

\begin{document}

\title{Sound propagation in  striped supersolid cold gases at zero temperature}

\author{Elena Poli}
\thanks{These authors contributed equally to this work.}
\affiliation{Pitaevskii BEC Center, CNR-INO and Dipartimento di Fisica, Universit\`a di Trento, I-38123 Trento, Italy}
\author{Giovanni I. Martone}
\thanks{These authors contributed equally to this work.}
\affiliation{CNR NANOTEC, Institute of Nanotechnology, Via Monteroni, 73100 Lecce, Italy}
\affiliation{INFN, Sezione di Lecce, 73100 Lecce, Italy}
\author{Sandro Stringari}
\affiliation{Pitaevskii BEC Center, CNR-INO and Dipartimento di Fisica, Universit\`a di Trento, I-38123 Trento, Italy}
\author{Alessio Recati}
\affiliation{Pitaevskii BEC Center, CNR-INO and Dipartimento di Fisica, Universit\`a di Trento, I-38123 Trento, Italy} 
\affiliation{Trento Institute for Fundamental Physics and Applications, INFN, I-38123 Trento, Italy}

\begin{abstract}
We present a unified hydrodynamic approach for the sound propagation in the stripe phases realized in ultracold dipolar gas and spin-orbit-coupled BEC platforms at zero temperature. Despite the deep difference of the two platforms at a microscopic level, a similar hydrodynamic description can be formulated at a macroscopic level. The main difference between the two platforms is the lack of Galilean invariance in the spin-orbit case, resulting in a different identification of the normal (nonsuperfluid) component of the density, which leads to new terms in the equation for the current. In both cases the spectrum comprises two sounds, reflecting the spontaneous breaking of the $U(1)$ and translational symmetries. Both sounds exhibit an anisotropic behavior. A comparison with the first and second sounds of the smectic-A liquid crystal is also presented. 
\end{abstract}

\maketitle

The recent experiments on the stripe phase of dipolar~\cite{he2025ooa,Laurianepvt} and spin-orbit-coupled (SOC)~\cite{Li2017,Putra2020,Chisholm2026} gases is expected to stimulate novel interest in the physics of superfluids exhibiting spontaneous breaking of translational symmetry, yielding supersolidity. 
Since the pioneering work by Andreev and Lifshitz~\cite{AndreevLifshitz}, it has been shown that in a $d$-dimensional crystalline structure in a $d$-dimensional space the presence of a finite superfluid fraction generates, according to the Goldstone theorem~\cite{Brauner2012}, $d+1$ sound modes. However, when the dimensionality of the system and the number of directions in which translational symmetry is spontaneously broken do not coincide, the counting and characterization of the resulting Goldstone modes become nontrivial. The stripe phase can be regarded as a one-dimensional crystal embedded in an effectively two-dimensional space, providing a concrete example of this subtle situation.

In this work, we will point out analogies and differences between the stripe phases in dipolar and spin-orbit-coupled gases by employing a hydrodynamic (HD) description. This framework naturally emphasizes the role of conservation laws and broken symmetries, and also allows for an insightful comparison with the hydrodynamic theory of the smectic phase of liquid crystals~\cite{Martin1972} and its superfluid extension, introduced in Ref.~\cite{Hofmann2021}.

A major difference between the microscopic Hamiltonian of a standard superfluid, including dipolar gases, and the SOC platforms, is the absence of full Galilean invariance in the latter, yet preserving translational invariance. This implies that even at zero temperature and in a homogeneous phase the superfluid density does not generally coincide with the total density~\cite{Zhang2016}. Instead, in the dipolar gas platform the superfluid density is smaller than the total density only in the nonuniform phase, where it essentially coincides with the Leggett bound for a weakly interacting Bose-Einstein condensate~\cite{Leggett1970}. 

We show that the absence of full Galilean invariance makes the HD equations for the current different on the two platforms.
Despite such a difference, in both platforms the homogeneous superfluid phase supports a single sound mode, namely the Nambu–Goldstone mode associated with the spontaneous breaking of the $U(1)$ symmetry. In the stripe phase an additional sound mode appears, corresponding to the breaking of translational invariance and associated with the crystalline order. As we will show, the nature of the sound modes depends strongly on the propagation direction with respect to the stripe orientation.

Let us point out that the stripes considered in the present work are due to the spontaneous breaking of translational invariance. They should not be confused with the stripes caused by the application of an external periodic potential, giving rise to artificial crystal structures. The latter configurations exhibit only a single Goldstone mode, associated with the spontaneous breaking of the $U(1)$ symmetry, and the corresponding gapless sound modes have already been the object of experimental investigation (see, for example, Ref.~\cite{Chauveau2023}). 

\begin{figure}
    \centering
    \includegraphics[width=0.9\linewidth]{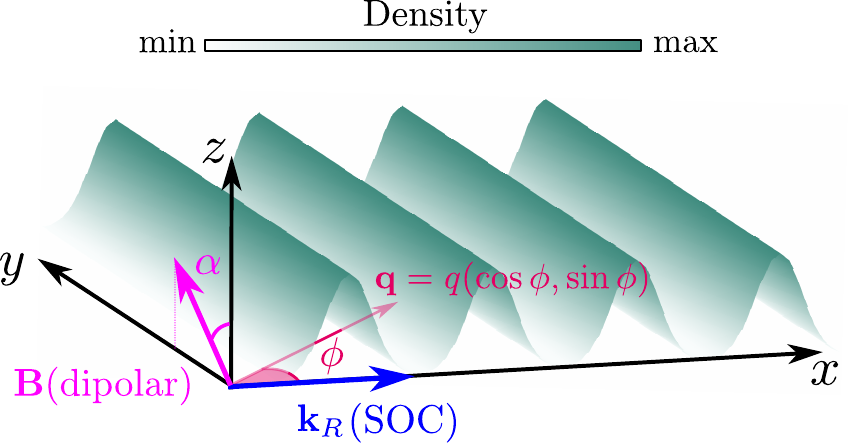}
    \caption{Illustrative sketch of a stripe supersolid in the $x-y$ plane. The blue arrow $\mb{k}_R$ indicates the direction of the Raman field for a spin–orbit-coupled system, while the magenta arrow $\mb{B}$ denotes the direction of the magnetic field for a dipolar system. The red arrow represents the generic wave vector $\mb{q}$ of a propagating sound wave. }
    \label{fig:stripe}
\end{figure}

In the following, we consider systems with stripes occurring in the $x-y$ plane. In both dipolar and spin-orbit platforms, the stripe orientation can be controlled by an external field. In dipolar gases, when the magnetic field is aligned along the $z$ direction, the stripe orientation is arbitrary~\cite{Ripley2023tds}. To fix it, we consider a tilted magnetic field, with the stripes aligned along its in-plane direction (see Fig.\,\ref{fig:stripe} and discussion below). In spin-orbit-coupled gases the stripe orientation is always perpendicular to the momentum imparted by the Raman laser fields that generate the spin-orbit coupling. In the following we assume the stripes to be oriented along the $y$ axis (see Fig.\,\ref{fig:stripe}).

\paragraph*{HD approach to the dipolar stripe phase.}
Supersolidity in ultracold dipolar gases originates from the interplay between short-range contact interactions and long-range dipolar interactions~\cite{F1,F2,S1,S2,I1,I2,Norcia2021_2D,Bland2021_2D}. These are described by the pseudopotential
\begin{equation}
V(\mb{r}) = \frac{4 \pi \hbar^2 a_s}{m} \delta(\mb{r}) + \frac{3 \hbar^2 a_{dd}}{m} \frac{1-3\,(\hat{\mb{d}}\cdot \mb{r})^2}{r^3} \, ,
\label{eq:dip_pseudopot}
\end{equation}
where $a_s$ is the tunable $s$-wave scattering length, $a_{dd}$ is the dipolar length, $m$ is the atomic mass, and $\hat{\mb{d}} = (0,\sin\alpha,\cos\alpha)$ is the dipole polarization direction set by a magnetic field $\mb{B}$ (see Fig.~\ref{fig:stripe}). In the planar geometry with the system harmonically confined along $z$, depending on the ratio $\epsilon_{dd} = a_{dd} / a_s$ and the average density $\rho$, different supersolid patterns can emerge~\cite{Ripley2023tds}. Here, we focus on the stripe phase, whose stability region in the phase diagram is enhanced by tilting the dipoles away from the vertical direction ($\alpha \neq 0$)~\cite{lima2025sdp}.

For the stripe phase of dipolar gases, the hydrodynamic theory is built in the standard way by considering particle-number and momentum conservation and the two order parameters related to the spontaneous breaking of the  $U(1)$ and Galilean symmetries, respectively. The long-wavelength, low-energy dynamics of the system occurs effectively in two dimensions. The equations for the 5 hydrodynamic variables, i.e. the average density $\rho$, the current $\mb{j}$, the gradient of the superfluid phase $\nabla \theta$, and the the gradient of stripe displacement $\nabla u$, read~\cite{Dorsey2010,ZawiAnomalous}
\begin{subequations}
\label{eqs:dip_hd_eqs}
\begin{eqnarray}
\p_t \rho + \p_i j^i = 0 \, ,
\label{eq:dip_n} \\
\p_t j^i + \p_j \Pi^{ij} = 0 \, ,
\label{eq:dip_j} \\
\p_t \mb{v}_\mathrm{s} + \nabla\mu = 0 \, ,
\label{eq:dip_vs} \\
\p_t \nabla u - \nabla v_\mathrm{n}^x = 0 \, ,
\label{eq:dip_u}
\end{eqnarray}
\end{subequations}
where Einstein summation notation is used, and we introduced the superfluid (normal) velocity $\mb{v}_\mathrm{s}=\hbar\nabla\theta/m$ ($\mb{v}_\mathrm{n}$), the stress tensor $\Pi^{ij}$, and the chemical potential per unit mass $\mu$. Equations~\eqref{eqs:dip_hd_eqs} can also be regarded as the HD equations for a superfluid smectic-A phase, as discussed in Ref.~\cite{Hofmann2021}.

We are interested in the long-wavelength, low-energy excitations, which can be obtained by linearizing the HD equations around the ground-state solution $(\rho_0,\mb{j}_0=\mb{0},\mb{v}_{\mathrm{s},0}=\mb{0},u_0=0)$. The linearized constitutive relations are given by
\begin{subequations}
\label{eqs:const_rels}
\begin{eqnarray}
\delta j^x & = & \rho_\mathrm{s} \delta v_\mathrm{s}^x + \rho_\mathrm{n} \delta v_\mathrm{n}^x \, ,
\label{eq:djx} \\
\delta j^y & = & \rho_0 \delta v_\mathrm{s}^y \, ,
\label{eq:djy} \\
\delta \Pi^{ij} & = & \kappa^{-1} \delta \rho\,\delta^{ij} - \lambda_x \p_x u \delta^{ix}\delta^{jx} - \lambda_y \p_y u \delta^{ix}\delta^{jy} \, ,
\label{eq:dp} \\
\delta \mu & = & (\rho_0\kappa)^{-1} \delta \rho \, ,
\label{eq:dmu}
\end{eqnarray}
\end{subequations}
where $\rho_\mathrm{s} < \rho_0$ and $\rho_\mathrm{n} = \rho_0 - \rho_\mathrm{s}$ are the superfluid and normal densities along $x$, respectively, $\kappa$ is the compressibility per unit mass, and $\lambda_{x(y)}$ are two elastic constants. The latter are the strengths of the restoring force due to the stripe displacement, $\lambda_x \p_x u$, and to the stripe rotation, $\lambda_y \p_y u$. When the magnetic field $\mathbf{B}$ is aligned along $z$, i.e., $\alpha=0$, there is no preferred orientation for the stripes in the $x-y$ plane, the choice of the stripe direction $x$ is arbitrary and there is no rotational restoring force. 
When instead $\alpha\neq 0$, the stripes have a preferred orientation and we expect $\lambda_y\neq 0$. We notice that a similar contribution in the stress tensor arises in the context of the smectic-A phase of liquid crystals in presence of an external magnetic field orthogonal to the layers~\cite{DeGennesBookLC}. Additionally, for the sake of simplicity, in the above equations we neglected any strain-density coupling, which has been shown to be very small for dipolar gases~\cite{Blakie2023sounds}, and does not modify the qualitative behavior of the sounds.  

We write the linearized Eqs.~\eqref{eqs:dip_hd_eqs} in momentum space and look for propagating solutions with wave vector $\mb{q}$ and frequency $\omega = c q$. We  introduce the longitudinal [$j_l = (j_x q_x + j_y q_y)/q$], and transverse [$j_t = (j_y q_x - j_x q_y)/q$], components of the current, to write the equations as:
\begin{subequations}
\label{eqs:dip_lin_eqs}
\begin{align}
&{} - c \delta\rho + \delta j_l = 0 \, ,
\label{eq:n_q} \\
&{} - c \delta j_l + \kappa^{-1} \delta\rho - \cos\phi \lambda_\phi \tilde u = 0 \, ,
\label{eq:j_l} \\
&{} - c \delta j_t + \sin\phi\lambda_\phi \tilde u = 0 \, ,
\label{eq:j_t} \\
&{} - c \delta v_\mathrm{s} + (\rho_0\kappa)^{-1} \delta \rho = 0 \, ,
\label{eq:vs_q} \\
&{} - c \tilde u - \delta j_l \frac{\cos\phi}{\rho_\mathrm{n}} + \delta j_t \frac{\sin\phi}{\rho_\mathrm{n}} + \delta v_\mathrm{s} \cos\phi \frac{\rho_\mathrm{s}}{\rho_\mathrm{n}} = 0 \, ,
\label{eq:u_q}
\end{align}
\end{subequations}
where we use the fact that the superfluid velocity is only longitudinal, we define $\tilde u=iqu$ and the angle-dependent elastic constant $\lambda_\phi=\lambda_x\cos^2\phi+\lambda_y\sin^2\phi$. 

The solutions are anisotropic phononic modes with speeds of sound $c(\phi)$, where $\phi=\arccos(\mb{q} \cdot \hat{\mb{x}}/q)$ is the angle between the propagation direction and the $x$ axis (see Fig.~\ref{fig:stripe}). 
The two finite sound speeds obtained from Eqs.~\eqref{eqs:dip_lin_eqs} can be written as
\begin{equation}
c_{1,2}^2(\phi)\!=\!\frac{1}{{2}}\left[\frac{1}{\kappa}+\frac{\lambda_\phi}{\rho_\mathrm{n}}\pm
\sqrt{\Big(\frac{1}{\kappa}+\frac{\lambda_\phi}{\rho_\mathrm{n}}\Big)^2\!\!-4 \frac{f_{\mathrm{s},\phi}}{\kappa}\frac{\lambda_\phi}{\rho_\mathrm{n}}}\right] \, ,
\label{eq:dip_sound_vel}
\end{equation}
where we define the superfluid fraction $ f_{\mathrm{s},\phi} = f_\mathrm{s}\cos^2\phi + \sin^2\phi$, with $f_\mathrm{s} = \rho_\mathrm{s}/\rho_0$ the superfluid fraction along $x$.

\begin{figure}
    \centering
    \includegraphics[width=0.9\linewidth]{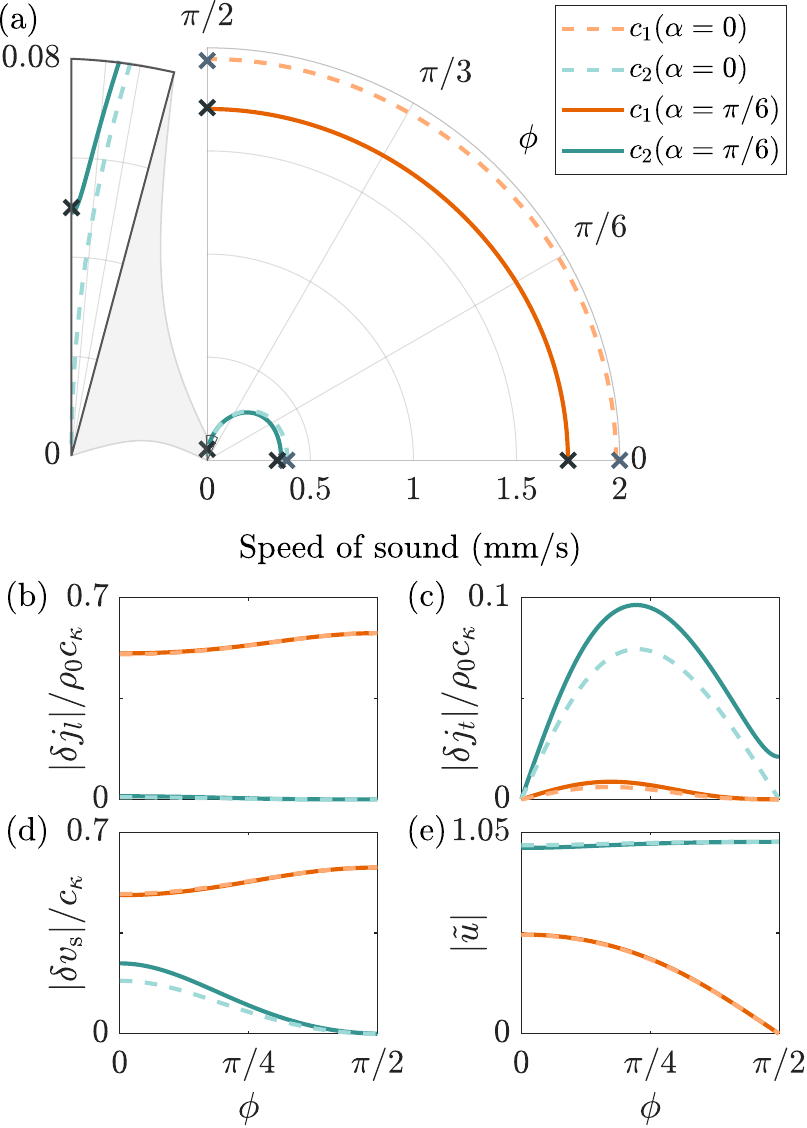}
    \caption{(a) Speeds of sound $c_{1,2}$ and (b)-(e) components of the eigenvectors associated with the hydrodynamic equations of a dipolar supersolid in the stripe phase for different propagation angles $\phi$, with ($\alpha=\pi/6$, solid line) and without ($\alpha=0$, dashed line) tilt of the magnetic field. The black and gray crosses are the speeds of sound obtained from real-time simulations with the eGPE (see End Matter). Hydrodynamic parameters for $\alpha = \pi/6$: $\rho_\mathrm{s}/\rho = 0.48$, $\kappa^{-1} = 2.9$ mm$^2$/s$^2$, $\lambda_x/\rho_\mathrm{n} = 0.28$ mm$^2$/s$^2$, $\lambda_y/\rho_\mathrm{n} = 0.0048$ mm$^2$/s$^2$. Hydrodynamic parameters for $\alpha = 0$: $\rho_\mathrm{s}/\rho = 0.54$, $\kappa^{-1} = 3.8$ mm$^2$/s$^2$, $\lambda_x/\rho_\mathrm{n} = 0.29$ mm$^2$/s$^2$, $\lambda_y/\rho_\mathrm{n} = 0$ mm$^2$/s$^2$. \textcolor{red}
    }
    \label{fig:dipolar-hd}
\end{figure}

In Fig.\,\ref{fig:dipolar-hd}(a), we report the speeds of sound as a function of the propagation angle $\phi$. 
The parameters $\rho_\mathrm{s(n)}$, $\kappa$, $\lambda_{x(y)}$ entering the HD equations are obtained by using the extended Gross-Pitaevskii equation (eGPE)~\cite{Wachter2016} for a dipolar gas of $^{164}$Dy atoms, harmonically confined along $z$ by a potential $V(z)=m\omega_z^2z^2/2$ with a frequency $\omega_z/(2\pi) = 72.4$\,Hz, a density $\rho_0/m = 2500\,\mu m^{-2}$ and $\epsilon_{dd} = 1.29$ (see End Matter). The extracted values are reported in the caption of Fig.\,\ref{fig:dipolar-hd}. In Fig.\,\ref{fig:dipolar-hd}(b)-(e), we show the components $\delta j_l$, $\delta j_t$, $\delta v_\mathrm{s}$, and $\tilde{u}$ of the eigenvectors of the HD equations, which allow us to characterize the sound modes in terms of their longitudinal or transverse, and superfluid or crystalline nature.

For $\phi=0$ we have two longitudinal propagating modes of mixed crystal and superfluid nature, as already discussed in literature for a one-dimensional supersolid structure \cite{Hofmann2021,Sindik2024,Platt2024}. For $\phi=\pi/2$, the equations for $(\delta\rho,\delta j_l,\delta v_\mathrm{s})$ and $(\delta j_t, \tilde u)$ decouple. The first triplet gives rise to a standard superfluid sound mode $c_1(\phi=\pi/2)=c_\kappa=\sqrt{\kappa^{-1}}$, and the second couple to a dispersive transverse mode $c_2(\phi=\pi/2)=\sqrt{\lambda_y/\rho_\mathrm{n}}$ \footnote{Notice that the speed of the transverse sound in an isotropic (super)solid has a similar expression, where $\lambda_y$ is replaced by the second Lamé coefficient~\cite{Dorsey2010,Poli2024eoa}.}. As already mention, the latter vanishes once the rotational symmetry of the Hamiltonian in the $x-y$ plane is restored ($\alpha=0$). In general, the nature of the modes strongly depends on the propagation angle $\phi$. Notice that in the regimes where the stripes are very rigid ($\lambda_\phi \kappa/\rho_\mathrm{n}\gg 1$), one gets $c_2(\phi)=c_\kappa\sqrt{f_{\mathrm{s},\phi}}$. This result is consistent with the sound of a standard BEC confined in a one-dimensional optical lattice, and for which the experimental confirmation has recently been obtained for $\phi=0$~\cite{Chauveau2023}.

It is useful to remind that also the standard smectic-A HD ($\rho_s=0$ and $\lambda_y=0$) predicts two sound modes: a slightly anisotropic and mainly longitudinal first sound of collisional nature, and the so-called second sound, derived by Martin, Parodi and Pershan~\cite{Martin1972}, called like this in analogy with the Helium nomenclature. The second sound, due to the coupling between the transverse current and the displacement of the stripes, is equivalent to the dispersive transverse mode of a crystal, and corresponds to the Goldstone mode due to the spontaneous breaking of translational invariance. The speed of sound of such a mode is very anisotropic and vanishes for sound propagation orthogonal or parallel to the stripes/layers. 

In our stripe system, superfluidity does not introduce a new mode, but modifies the smectic second sound. In particular, for $\phi = 0$, the coupling to the superfluid velocity makes the so-called permeation mode dispersive rather than diffusive~\cite{Hofmann2021}, making the system equivalent to a one-dimensional supersolid. Instead, for any $\phi \neq 0$ the superfluidity just modifies the transverse mode already present in the smectic description. Concerning the first sound, although its structure appears similar to that of a conventional smectic-A phase, its physical origin is different: the collisional hydrodynamic sound of a standard liquid crystal is replaced here by a collisionless sound.

\paragraph*{HD approach to the SO stripe phase.} The hydrodynamic formulation of the stripe phase of a two-component Bose gas with spin-orbit coupling (SOC) is more subtle to build due to the spin nature of the system and the important modification of the single-particle Hamiltonian, for which we consider the form~\cite{Lin2011}
\begin{equation}
h_\mathrm{SO}= \frac{1}{2m} \left[(p_x-\hbar k_R\sigma_z)^2 + p^2_y\right]+\frac{\hbar\Omega_R}{2}\sigma_x+\frac{\hbar\delta_R}{2} \sigma_z \, ,
\label{eq:soc_ham}
\end{equation}
where $k_R$ is the modulus of the wave vector mismatch between the laser fields, $\Omega_R$ the Raman coupling fixed by their intensity, $\delta_R$ an effective Raman detuning, and $\sigma_{x,z}$ are the Pauli matrices. The two-body interaction is described by the spin-dependent zero-range pseudopotential typical of a spin mixture~\cite{BecBook2016}, $V_{\sigma\sigma'}(\mb{r}) = g_{\sigma\sigma'} \delta(\mb{r})$ ($\sigma,\sigma' = \uparrow,\downarrow$), where $g_{\sigma\sigma'} = 4\pi \hbar^2 a_{s,\sigma\sigma'}/m$ and $a_{s,\sigma\sigma'}$ are the scattering lengths in the various spin channels. The ground-state and dynamic properties of these systems have been extensively studied (see the reviews in~\cite{Zhai2015review,Zhang2016review,Martone2025review} and references therein). A peculiarity of the SOC Hamiltonian is that it breaks explicitly Galilean invariance since the physical momentum operator $p_x-\hbar k_R \sigma_z$ does not commute with the Raman term $\hbar\Omega_R\sigma_x/2$. However, $h_\mathrm{SO}$ commutes with the canonical momentum $p_x$ and thus enjoys translational invariance. The latter is a crucial difference with respect to the more standard case of a superfluid in which the continuous translational symmetry is broken due to periodic or disorder potentials, like, e.g., gases in optical lattices or $^4$He in porous media.

We start by representing the order parameter of the condensate in the supersolid configuration as a coherent mixture of two different momentum states, $\Psi(\mb{r},t) = \sum_{\ell = \pm} \sqrt{n_\ell} \Phi_\ell e^{i \theta_\ell(\mb{r},t)}$. Here $\Phi_\pm$ are the two-component real spinor wave functions of the two states, taken normalized to $1$, and $\theta_\pm(\mb{r},t) = \theta_{o,\pm} + k_\pm x - \mu_\pm t / \hbar$ their phases~\cite{Martone2026a}. Up to corrections due to interactions, the condensation momenta $\hbar k_+ > 0$ and $\hbar k_- < 0$ match the two minima of the spectrum of the single-particle Hamiltonian~\eqref{eq:soc_ham}. The chemical potentials $\mu_\pm$ fix the individual densities $n_\pm$ of the two momentum components, while the phase constants $\theta_{o,\pm}$ are arbitrary, reflecting the presence of two spontaneously broken symmetries. Since the spinors $\Phi_\pm$ are not generally orthogonal to each other, the microscopic density profile $n(\mb{r},t) = \Psi^\dagger(\mb{r},t)\Psi(\mb{r},t)$ exhibits periodic modulations (stripes) with wave vector $2 k_1 = k_+ - k_-$. The equilibrium condition at fixed total density $\rho = m (n_+ + n_-)$ requires $\mu_+ = \mu_-$, corresponding to stationary stripes. When $\mu_+ \neq \mu_-$ the fringes move along $x$ at velocity $v_\mathrm{fr} = (\mu_+ - \mu_-) / 2 \hbar k_1$, and their displacement reads $u = - (\theta_{o,+} - \theta_{o,-}) / 2 k_1 + v_\mathrm{fr} t$. In the hydrodynamic regime, the relevant degrees of freedom are the total density $\rho$, the canonical momentum density $\mb{g} = \hbar (n_+ \nabla\theta_+ + n_- \nabla\theta_-)$, the velocity field $\mb{v}_\mathrm{s} = \hbar \nabla(\theta_+ + \theta_-) / 2 m$ \footnote{In general, the superflow velocity is given by the deviation of $\mb{v}_\mathrm{s}$ from its ground-state value $\hbar (k_+ + k_-) \hat{\mb{x}} / 2 m$. In the $\mathbb{Z}_2$ case studied in the following the ground state value is zero.}, and the gradient of the stripe displacement $\nabla u$, all taken to vary slowly in space and time. The time evolution of these variables is related to the chemical potential per unit mass $\mu = (\mu_+ + \mu_-) / 2 m$, the current $\mb{j}$, the stress tensor $\Pi^{ij}$, and the fringe velocity $v_\mathrm{fr}$ according to \footnote{Although all the quantities have been introduced using a two-harmonic approximation for $\Psi$, generalizing their definitions to include higher-order terms induced by the interactions is straightforward~\cite{Martone2026a,Martone2026b}.}
\begin{subequations}
\label{eqs:soc_hd_eqs}
\begin{eqnarray}
\p_t \rho + \p_i j^i = 0 \, ,
\label{eq:soc_n} \\
\p_t g^i + \p_j \Pi^{ij} = 0 \, ,
\label{eq:soc_g} \\
\p_t \mb{v}_\mathrm{s} + \nabla \mu = 0 \, ,
\label{eq:soc_vth} \\
\p_t \nabla u - \nabla v_\mathrm{fr} = 0 \, .
\label{eq:soc_u}
\end{eqnarray}
\end{subequations}
These hydrodynamic equations can be obtained directly by coarse graining the Gross-Pitaevskii equations. The procedure is however rather involved and will be described in another work~\cite{Martone2026b}. 

Note that the breaking of Galilean invariance introduces important differences compared to the dipolar case. First, the canonical momentum $g^x$, and not the physical momentum $j^x$, is conserved in the SO platform (parallel to the stripes one has $g^y = j^y$). Another difference is that we cannot identify the velocity $v_\mathrm{fr}$ in Eq.~\eqref{eq:soc_u} with the normal component velocity, as we did for the dipolar case [see, in particular, Eq.~\eqref{eq:dip_u}]. This follows from the proper identification of the normal (nonsuperfluid) density (see discussion below).

\begin{figure}
    \centering
    \includegraphics[width=0.9\linewidth]{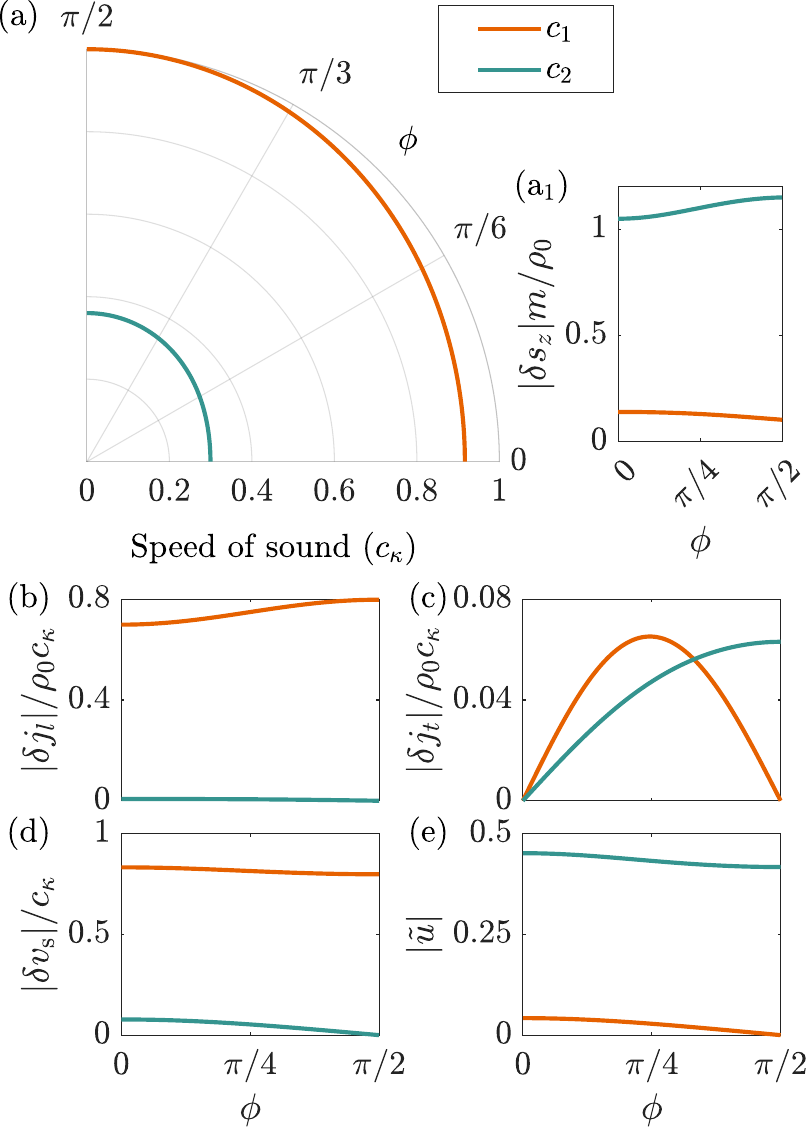}
    \caption{(a) Speeds of sound $c_{1,2}$, ($a_1$) spin polarization, and (b)-(e) components of the eigenvectors associated with the hydrodynamic equations of a SOC supersolid in the stripe phase for different propagation angles $\phi$. Hydrodynamic parameters: $\rho_\mathrm{s} / \rho_0 = 0.82$, $\rho_\mathrm{fr} / \rho_0 = 0.42$, $\kappa^{-1} / m E_R = 1.23$, $\gamma = 24.2$, $\Lambda_x / (\rho_\mathrm{fr} E_R / m) = 0.11$, $\Lambda_y / (\rho_\mathrm{fr} E_R / m) = 0.16$.}   
    \label{fig:SOC-hd}
\end{figure}

For the purpose of the present work and in order to have a clean comparison between the SOC and the dipolar stripe phases, we consider the easiest case where the microscopic model has a $\mathbb{Z}_2$ symmetry, i.e., $\delta_R = 0$ and $g_{\uparrow\uparrow} = g_{\downarrow\downarrow} = \bar{g}$. We then linearize around a ground state with parameters $\rho_0$, $\mb{g}_0 = \mb{0}$, $\mb{v}_{s,0} = \mb{0}$, and $u_0 = 0$. Neglecting again the lattice strain-density coupling, this yields the same constituent relations as for dipolar stripes, Eqs.~\eqref{eqs:const_rels}, with the current along $x$ modified as follows:
\begin{equation}
\delta j^x = \rho_\mathrm{s} \delta v_\mathrm{s}^x + \rho_\mathrm{fr} \delta v_\mathrm{fr} + \rho_\mathrm{laser} \delta v_\mathrm{laser} \, .
\label{eq:soc_djx_laser}
\end{equation}
The parameter $\rho_\mathrm{s}$ can be identified with the superfluid density along $x$, and can be proven to coincide with the value obtained as the difference between the total density and the static transverse current response function~\cite{baym1968,scalapino92}. The fringe density $\rho_\mathrm{fr}$ is the part of the density which participates in the motion of the fringes. Due to the absence of Galilean invariance at the level of the single-particle Hamiltonian, however, $\rho_\mathrm{s} + \rho_\mathrm{fr} \neq \rho_0$ and there is a part of the density which is locked to the Raman lasers and does not participate in the collective  motion. This part, that we denote as $\rho_\mathrm{laser}$, is in general nonzero even in the uniform phases of the system~\cite{Zhang2016}. It can be calculated by forcing the Raman lasers to  move along $x$ at a small velocity $\delta v_\mathrm{laser}$ and calculating the corresponding change $\rho_\mathrm{laser} \delta v_\mathrm{laser}$ in the current. One can prove that $\rho_\mathrm{fr}$ and $\rho_\mathrm{laser}$ coincide with the  
contributions to the static response to a transverse current, given by the changes of the canonical momentum and of the  spin terms, respectively (see End Matter). This yields~\cite{baym1968,scalapino92}
\begin{equation}
\rho_\mathrm{n} = \rho_\mathrm{fr} + \rho_\mathrm{laser}
\label{eq:soc_rho_n}
\end{equation} 
and ensures the identity $\rho_\mathrm{s} + \rho_\mathrm{fr} + \rho_\mathrm{laser} = \rho_0$. Remarkably, we find that in the supersolid phase of the SOC gas the value of $\rho_\mathrm{laser}$ is negative. By contrast, in the uniform phases $\rho_\mathrm{laser}$ coincides with $\rho_\mathrm{n}$ and is therefore positive~\cite{Zhang2016}. Thus, the SOC stripe phase can be seen as a zero-temperature analog of the three-fluid model for a supersolid recently suggested by Saslow~\cite{Saslow2025}. In our case, the normal component is not due to thermal fluctuations, but to the absence of Galilean invariance in the Hamiltonian~\eqref{eq:soc_ham}.

In addition to Eq.~\eqref{eq:soc_djx_laser}, we need the relation $\delta g^x = \delta j^x + \hbar k_R \delta s_z$ linking the canonical and physical momentum through the spin polarization term $\hbar k_R\delta s_z = \rho_\mathrm{laser} \delta v_\mathrm{s}^x + (\gamma-1) \rho_\mathrm{fr} \delta v_\mathrm{fr}$, with $\gamma$ a dimensionless HD parameter. Since $\delta v_\mathrm{fr}$ determines the population fluctuations in the momentum minima, and the latter is related to the spin polarization of the gas, the second term in $\delta s_z$ is expected to be related to the magnetic susceptibility $\chi$ of the system. In particular, for $\Omega_R \rightarrow 0$, we find that $(\gamma-1)\rho_\mathrm{fr}/\rho_0=2 k_1 E_R \chi/k_R$. The latter equality approximately holds as long as we are not too close to the transition to the uniform phase. The linearization of Eqs.~\eqref{eqs:soc_hd_eqs} in momentum space leads to expressions identical to Eqs.~\eqref{eqs:dip_lin_eqs}, with $\rho_\mathrm{fr}$ replacing $\rho_\mathrm{n}$ and Eqs.~\eqref{eq:j_l}-\eqref{eq:j_t} modified as follows:
\begin{subequations}
\label{eqs:soc_lin_eqs}
\begin{align}
&{} - c \delta j_l + \frac{\delta\rho}{\kappa_l(\phi)} - \cos\phi \Lambda_\phi \tilde u = 0 \, ,
\label{eq:lin_soc_eq_jl} \\
&{} - c \delta j_t + \frac{\delta \rho}{\kappa_t(\phi)} + \sin\phi \Lambda_\phi \tilde u = 0 \, ,
\label{eq:lin_soc_eq_jt}
\end{align}
\end{subequations}
where we define $\Lambda_\phi=\Lambda_x\cos^2\phi+\Lambda_y\sin^2\phi$ and, compared to the dipolar stripe phase, we have renormalized elastic constants $\Lambda_{x(y)} = \lambda_{x(y)} / \gamma$. More importantly, we have identified an angle-dependent longitudinal [$\kappa_l^{-1}(\phi)=\kappa^{-1}(\nu\cos^2\phi+\sin^2\phi)$] and transverse [$\kappa_t^{-1}(\phi)=\kappa^{-1}\sin\phi\cos\phi(1-\nu)$] effective compressibility, which takes into account the mass renormalization via the parameter $\nu = (\rho_s+\rho_\mathrm{fr}/\gamma) / \rho_0$.  Since $\kappa_t^{-1}=0$ for propagation along to the stripes, analogously to the dipolar gas case, one has a density superfluid sound mode with $c_1(\phi = \pi / 2) = \sqrt{\kappa^{-1}}$ and a transverse mode with $c_2(\phi = \pi / 2) = \sqrt{\Lambda_y/\rho_\mathrm{fr}}$.

For general $\phi$, one finds the two sound velocities
\begin{eqnarray}
c_{1,2}^2(\phi)\!=\!\frac{1}{{2}}\!\!\left[\!\frac{1}{\kappa_{l,\phi}}\!+\!\frac{\Lambda_\phi}{\rho_\mathrm{fr}}\pm\!
\sqrt{\Big(\frac{1}{\kappa_{l,\phi}}\!+\!\frac{\Lambda_\phi}{\rho_\mathrm{fr}}\Big)^2\!\!-4 \frac{f_{\mathrm{s},\phi}}{\kappa}\frac{\Lambda_\phi}{\rho_\mathrm{fr}}}\right]\!\!,
\label{eq:soc_sound_vel}
\end{eqnarray}
where again $f_{\mathrm{s},\phi} =\rho_\mathrm{s}/\rho_0 \cos^2 \phi + \sin^2\phi$. The speeds of sound for the stripe phases in the dipolar and SOC platforms have almost an identical expression, but for the angular dependence of the effective compressibility (absent in the term proportional to $f_{\mathrm{s},\phi}$). Due to the Raman lasers, $\Lambda_x$ and $\Lambda_y$ are similar in SOC platforms, making the second sound less anisotropic compared to the dipolar case. In the absence of stripes, corresponding to a vanishing value of $\Lambda_\phi/\rho_{\mathrm{fr}}$, we recover the expression for the only sound exhibited by  the uniform nonsupersolid single-minimum phase, $c_\mathrm{SM}(\phi) = \sqrt{f_\mathrm{s,\phi}/\kappa}$~\cite{Martone2012}.

For the SO platform the HD parameters $\rho_{\mathrm{s}(\mathrm{fr})}$, $\kappa$, $\gamma$, and $\lambda_{x(y)}$ can be calculated semianalytically~\cite{Martone2026b}. We use the values $\hbar\Omega_R / E_R = 2.0$, $\bar{g} \rho_0 / m E_R = 1.52$, and $g_{\uparrow\downarrow} \rho_0 / m E_R = 0.88$, with $E_R = \hbar^2 k_R^2 / 2 m$. The speeds of sound and the most relevant components of the eigenvectors are reported in Fig.~\ref{fig:SOC-hd}, where the values of the HD parameters are reported for completeness. Furthermore, in Fig.~\ref{fig:SOC-hd}$(a_1)$ we report the relevant results for $\delta s_z$ which explicitly show that, differently from the first sound, the second sound mode is characterized by sizable spin oscillating effects, as first pointed out in Ref.~\cite{Li2013}.

Concerning the nature of the modes, we notice that the changes with respect to the dipolar case are mostly quantitative, the main difference being in the transversality of the two sounds of Fig.~\ref{fig:SOC-hd}(c), due to the fact that the value of the rotational restoring force $\Lambda_y$ is comparable with the compressional elastic constant $\Lambda_x$, leading to a strong coupling between $\tilde u$ and $\delta j_t$ even for $\phi=\pi/2$. In systems with two-dimensional SOC, $\Lambda_y$ depends on the degree of SOC anisotropy and vanishes in the isotropic limit~\cite{Jian2011}.

\paragraph*{Conclusion.}
We have developed a unified hydrodynamic formalism to describe the anisotropic  propagation of sound in two-dimensional-like striped supersolid gases, including the case of dipolar and spin-orbit-coupled platforms at zero temperature. Our analysis has pointed out the crucial role played by the violation of Galilean invariance in the SOC case. The hydrodynamic parameters entering the relevant equations for the sound velocities have been calculated employing the Gross-Pitaevskii theory and  its extended version in the dipolar case. The transverse nature of the second sound mode propagating parallel to the stripes has been found to be particularly pronounced in the SOC platform, due to the large value of the rotational elastic restoring force, opening realistic perspectives of experimental observation (see also~\cite{Geier2023}). Future perspectives concern the propagation of sound at finite temperature, where the behavior of the temperature depletion of the superfluid density in nonuniform configurations is still under debate~\cite{hegg2024,berger2025} and one expects the occurrence of an additional propagating mode~\cite{Hofmann2021}. 

\paragraph*{Note added:} While completing the present work we became aware of a work on the sound in the supersolid phase of dipolar gases in the presence of a magnetic field not aligned with the $z$ direction. The work briefly discusses the stripe phase, reporting results in perfect agreement with our findings for the dipolar platform. In particular, it confirms the presence (and the small magnitude) of the restoring force for stripe rotation~\cite{Cook2026}.

\paragraph*{Acknowledgments.}
Support from the Provincia autonoma di Trento and from INFN through the RELAQS project is acknowledged. This work was supported by the Italian Ministry of University and Research (MUR) through the PNRR MUR project: `National Quantum Science and Technology Institute' - NQSTI (PE0000023) and the PNRR MUR project: `Integrated Infrastructure Initiative in Photonic and Quantum Sciences' - I-PHOQS (IR0000016). We acknowledge the support of the Quantum Optical Networks based on Exciton-polaritons - (Q-ONE) funding from the HORIZON-EIC-2022-PATHFINDER CHALLENGES EU programme under grant agreement No. 101115575, and of the Neuromorphic Polariton Accelerator - (PolArt) funding from the Horizon-EIC-2023-Pathfinder Open EU programme under grant agreement No. 101130304. Views and opinions expressed are however those of the author(s) only and do not necessarily reflect those of the European Union or European Innovation Council and SMEs Executive Agency (EISMEA). Neither the European Union nor the granting authority can be held responsible for them.

\newpage
\begin{center}
\textbf{\large End Matter}
\end{center}
\section*{Numerical details for dipolar systems}
While for the SOC systems the hydrodynamic elastic parameters can be derived semianalytically, for the dipolar case we rely on numerical approaches based on the Gross-Pitaevskii equation\,\cite{Wachter2016}. The extended Gross-Pitaevskii equation (eGPE) for a dipolar system with condensate wave function $\Psi$ reads
\begin{equation}\label{epge}
\begin{aligned}
i \hbar \frac{\partial \Psi(\mathbf{r}, t)}{\partial t} = & {\left[-\frac{\hbar^2 \nabla^2}{2 m}\right.} + V_{trap}\\
& +\int \text{d}^3\mathbf{r}^{\prime}\, V\left(\mathbf{r}-\mathbf{r}^{\prime}\right)\left|\Psi\left(\mathbf{r}^{\prime}, t\right)\right|^2  \\
& +\gamma_{\mathrm{QF}}|\Psi(\mathbf{r}, t)|^3\biggr]  \Psi(\mathbf{r}, t)\,,
\end{aligned}
\end{equation}
where the interaction pseudopotential $V(\mathbf{r})$ is reported in the main text, $V_{trap} = \frac{1}{2}m\omega_z^2z^2$ is an harmonic confinement along $z$ and $\gamma_{\mathrm{QF}}=\frac{128 \hbar^2}{3 m} \sqrt{a_{\mathrm{s}}^5} \operatorname{Re}\left\{\mathcal{Q}_5\left(\epsilon_{\mathrm{dd}}\right)\right\}$ is the coefficient of the Lee-Huang-Yang term describing quantum fluctuations \cite{FischerLHY,Pelster}, with $\mathcal{Q}_5\left(\varepsilon_{\mathrm{dd}}\right)=\int_0^1 d u\left(1-\varepsilon_{\mathrm{dd}}+3 u^2 \varepsilon_{\mathrm{dd}}\right)^{5 / 2}$. The latter is necessary for the stabilization of the supersolid state against collapse\,\cite{Bisset2016gsp}. 

Since the system is infinite in the $x-y$ plane, for the analysis of ground-state properties we restrict the study of the wave function to a single unit cell in the plane. To study the supersolid stripe phase we adopt a square unit cell with lattice constant $a$ and impose periodic boundary conditions. For a fixed value of the density $\rho_0$, we then compute the minimum energy solution for different values of the lattice constant $a$, and determine its optimal value $a^*$ that minimizes the energy density $\varepsilon$. 

The superfluid fraction of the ground-state configuration is evaluated using the Leggett formula\,\cite{Leggett1970}. The other elastic parameters $\kappa,\lambda_x,\lambda_y$ are obtained by introducing small perturbations around the ground-state configuration, recalculating the corresponding energies, and extracting the coefficients from the resulting energy variations. 
More in details, we calculate
\begin{subequations}
\begin{eqnarray}
    \kappa^{-1} = \frac{1}{m}\rho_0\frac{\partial^2 \varepsilon}{\partial \rho^2}\,, \\
    \lambda_x = \frac{1}{m\rho_\mathrm{n}}(a^{*})^2 \frac{\partial^2 \varepsilon}{\partial a^2}\,, \\
    \lambda_y = \frac{1}{m\rho_\mathrm{n}} \frac{\partial^2 \varepsilon}{\partial \beta^2}\,,
\end{eqnarray}
\end{subequations}
where $\beta$ is the polar angle in the $x-y$ plane of the magnetic field $\mathbf{B}$ with respect to the $y$ axis (in Fig.\,\ref{fig:stripe} the magnetic field vector is along $y$, so $\beta = 0$).

To benchmark the hydrodynamic predictions, we perform real-time simulations of the extended Gross–Pitaevskii equation (eGPE). In these simulations, long-wavelength perturbations are applied, and the resulting excitation frequencies are extracted from the density response. To excite the two sound modes at $\phi = 0$ and the first sound mode at $\phi = \pi/2$, we replicate the unit cell 50 times along either the $x$ or $y$ direction. The initial state at $t = 0$ is then constructed by imprinting a longitudinal phase modulation onto the ground-state wavefunction
\begin{equation}
    \Psi(\mb{r},t=0) = \Psi_{GS}(\mb{r})\,e^{i \delta \sin(k_x x + k_y y)}\,.
\end{equation}
Here, we set $(k_x,k_y) = (2\pi/50a^*,0)$ for $\phi=0$,  $(k_x,k_y) = (0,2\pi/50a^*)$ for $\phi=\pi/2$, and $\delta \ll 1$. The system is then evolved up to a total time $T = 5\,\mathrm{s}$, while monitoring the density response function
\begin{equation}
R(t) = \int d^3 \mb{r} \sin(k_x x + k_y y) |\Psi(\mb{r},t)|^2
\end{equation}
By Fourier transforming the resulting oscillating signal $R(t)$, we extract the excitation frequencies from the spectral peaks. These are shown as black and gray crosses in Fig.~\ref{fig:dipolar-hd} of the main text, and exhibit excellent agreement with the hydrodynamic predictions.

Importantly, for the excitation of the second sound mode at $\phi = \pi/2$, a different phase perturbation is required, as this mode is purely transverse, as seen in Fig.\,\ref{fig:dipolar-hd}\,(b)-(c). In this case, we initialize the system as
\begin{equation}
    \Psi(\mb{r},t=0) = \Psi_{GS}(\mb{r})\,e^{i \delta_1 \sin(k_y y + \delta_2\sin k_x x)}\,,
\end{equation}
with $\delta_1, \delta_2 \ll 1$. The corresponding response function is defined as
\begin{equation}
    R(t) = \int d^3 \mb{r} \sin(k_y y + \delta_2\sin k_x x) |\Psi(\mb{r},t)|^2\,.
\end{equation}
Also in this case, the extracted frequency is in excellent agreement with the hydrodynamic results.

\section*{Normal density in spin-orbit-coupled Bose gases}
A common method to directly access the normal density of a many-body system consists in the calculation of the static response to a transverse current perturbation~\cite{baym1968,scalapino92}. In spin-orbit-coupled Bose gases, the transverse current operator along $x$ reads~\cite{Zhang2016}
\begin{equation}
J_x^T(\mb{q}) = P_x(\mb{q}) - \hbar k_R \Sigma_z(\mb{q}) \, ,
\label{eq:soc_trans_curr}
\end{equation}
where $\mb{q} = q_y \hat{\mb{y}}$ lies along $y$ and we have introduced the $\mb{q}$ component of the canonical momentum density,
\begin{equation}
P_x(\mb{q}) = \sum_{k=1}^N p_{x,k} e^{- i q_y y_k} \, ,
\label{eq:soc_can_curr_op}
\end{equation}
and of the spin density,
\begin{equation}
\Sigma_z(\mb{q}) = \sum_{k=1}^N \sigma_{z,k} e^{- i q_y y_k} \, .
\label{eq:soc_spin_dens_op}
\end{equation}
The sums in the last two equations run over the $N$ atoms in the gas. Let $| 0 \rangle$ be the system's ground state, taken in the stripe phase, and $|b,\mb{q}\rangle$ its excited states, each identified by the index $b$ of the excitation band it belongs to and by its quasimomentum $\mb{q}$. The corresponding energies are denoted by $E_0$ and $E_{b,\mb{q}}$, respectively. Then, the normal fraction along $x$ is given by~\cite{baym1968,scalapino92}
\begin{equation}
\frac{\rho_\mathrm{n}}{\rho_0} = \frac{1}{N} \lim_{\mb{q} \to 0} \sum_b \left[ \frac{|\langle 0| J_x^T(\mb{q}) |b,\mb{q}\rangle|^2}{E_{b,\mb{q}} - E_0} + (\mb{q} \to - \mb{q}) \right] \, .
\label{eq:soc_trans_curr_rho_n}
\end{equation}
To recover the decomposition~\eqref{eq:soc_rho_n} of $\rho_\mathrm{n}$, we note that $\rho_\mathrm{fr}$ can be regarded as the response of the current to a translation motion of the fringes. We recall that the translations of the stripe wave function are generated by the canonical momentum, i.e., the $\mb{q} \to 0$ limit of the operator $P_x(\mb{q})$. We thus write
\begin{equation}
\begin{aligned}
\frac{\rho_\mathrm{fr}}{\rho_0} = \frac{1}{N} \lim_{\mb{q} \to 0} \sum_b & \Bigg[ \frac{\langle 0| J_x^T(\mb{q}) |b,\mb{q}\rangle \langle 0| P_x(\mb{q}) |b,\mb{q}\rangle}{E_{b,\mb{q}} - E_0} \\ & + (\mb{q} \to - \mb{q}) \Bigg] \, .
\label{eq:soc_trans_curr_rho_fr}
\end{aligned}
\end{equation}
To deduce $\rho_\mathrm{laser}$, we take the Raman lasers to move at a small velocity $\delta v_\mathrm{laser}$. This motion induces a Doppler shift of the effective frequency of the laser field felt by the atoms, resulting in the addition of a detuning term $\hbar k_R \sigma_z \delta v_\mathrm{laser}$ to the single particle Hamiltonian~\eqref{eq:soc_ham}. Noting that $\Sigma_z(\mb{q})$ reduces to the total spin operator along $z$ as $\mb{q} \to 0$, the response of the current to the laser motion is given by
\begin{equation}
\begin{aligned}
\frac{\rho_\mathrm{laser}}{\rho_0} = \frac{- \hbar k_R}{N} \lim_{\mb{q} \to 0} \sum_b & \Bigg[ \frac{\langle 0| J_x^T(\mb{q}) |b,\mb{q}\rangle \langle 0| \Sigma_z(\mb{q}) |b,\mb{q}\rangle}{E_{b,\mb{q}} - E_0} \\ 
& + (\mb{q} \to - \mb{q}) \Bigg] \, .
\label{eq:soc_trans_curr_rho_laser}
\end{aligned}
\end{equation}
Note that summing the right-hand sides of Eqs.~\eqref{eq:soc_trans_curr_rho_fr} and~\eqref{eq:soc_trans_curr_rho_laser} one recovers the full current response~\eqref{eq:soc_trans_curr_rho_n}. All these quantities can be easily evaluated using the Bogoliubov approach~\cite{Li2013,Martone2021}. We have checked that the obtained values of $\rho_\mathrm{fr}$ and $\rho_\mathrm{laser}$ are in agreement with those computed within the hydrodynamic theory. In particular, we find that the laser contribution $\rho_\mathrm{laser}$ to the normal density is sizable and negative, compensating the large and positive contribution given by $\rho_\mathrm{fr}$. This differs from the case of uniform SOC configurations, where $\rho_\mathrm{fr} = 0$ due to the absence of stripes and $\rho_\mathrm{laser} \equiv \rho_\mathrm{n}$ is positive~\cite{Zhang2016}.


\begin{thebibliography}{54}%
	\makeatletter
	\providecommand \@ifxundefined [1]{%
		\@ifx{#1\undefined}
	}%
	\providecommand \@ifnum [1]{%
		\ifnum #1\expandafter \@firstoftwo
		\else \expandafter \@secondoftwo
		\fi
	}%
	\providecommand \@ifx [1]{%
		\ifx #1\expandafter \@firstoftwo
		\else \expandafter \@secondoftwo
		\fi
	}%
	\providecommand \natexlab [1]{#1}%
	\providecommand \enquote  [1]{``#1''}%
	\providecommand \bibnamefont  [1]{#1}%
	\providecommand \bibfnamefont [1]{#1}%
	\providecommand \citenamefont [1]{#1}%
	\providecommand \href@noop [0]{\@secondoftwo}%
	\providecommand \href [0]{\begingroup \@sanitize@url \@href}%
	\providecommand \@href[1]{\@@startlink{#1}\@@href}%
	\providecommand \@@href[1]{\endgroup#1\@@endlink}%
	\providecommand \@sanitize@url [0]{\catcode `\\12\catcode `\$12\catcode
		`\&12\catcode `\#12\catcode `\^12\catcode `\_12\catcode `\%12\relax}%
	\providecommand \@@startlink[1]{}%
	\providecommand \@@endlink[0]{}%
	\providecommand \url  [0]{\begingroup\@sanitize@url \@url }%
	\providecommand \@url [1]{\endgroup\@href {#1}{\urlprefix }}%
	\providecommand \urlprefix  [0]{URL }%
	\providecommand \Eprint [0]{\href }%
	\providecommand \doibase [0]{https://doi.org/}%
	\providecommand \selectlanguage [0]{\@gobble}%
	\providecommand \bibinfo  [0]{\@secondoftwo}%
	\providecommand \bibfield  [0]{\@secondoftwo}%
	\providecommand \translation [1]{[#1]}%
	\providecommand \BibitemOpen [0]{}%
	\providecommand \bibitemStop [0]{}%
	\providecommand \bibitemNoStop [0]{.\EOS\space}%
	\providecommand \EOS [0]{\spacefactor3000\relax}%
	\providecommand \BibitemShut  [1]{\csname bibitem#1\endcsname}%
	\let\auto@bib@innerbib\@empty
	\bibitem [{\citenamefont {He}\ \emph {et~al.}(2025)\citenamefont {He},
		\citenamefont {Zhen}, \citenamefont {Parit}, \citenamefont {Huang},
		\citenamefont {Defenu}, \citenamefont {Boronat}, \citenamefont
		{S{\'a}nchez-Baena},\ and\ \citenamefont {Jo}}]{he2025ooa}%
	\BibitemOpen
	\bibfield  {author} {\bibinfo {author} {\bibfnamefont {Y.}~\bibnamefont
			{He}}, \bibinfo {author} {\bibfnamefont {H.}~\bibnamefont {Zhen}}, \bibinfo
		{author} {\bibfnamefont {M.~K.}\ \bibnamefont {Parit}}, \bibinfo {author}
		{\bibfnamefont {M.}~\bibnamefont {Huang}}, \bibinfo {author} {\bibfnamefont
			{N.}~\bibnamefont {Defenu}}, \bibinfo {author} {\bibfnamefont
			{J.}~\bibnamefont {Boronat}}, \bibinfo {author} {\bibfnamefont
			{J.}~\bibnamefont {S{\'a}nchez-Baena}},\ and\ \bibinfo {author}
		{\bibfnamefont {G.-B.}\ \bibnamefont {Jo}},\ }\bibfield  {title} {\bibinfo
		{title} {{Observation of a supersolid stripe state in two-dimensional dipolar
				gases}},\ }\href {https://arxiv.org/abs/2512.13280} {\bibfield  {journal}
		{\bibinfo  {journal} {arXiv:2512.13280 [cond-mat.quant-gas]}\ } (\bibinfo
		{year} {2025})}\BibitemShut {NoStop}%
	\bibitem [{\citenamefont {Chomaz}()}]{Laurianepvt}%
	\BibitemOpen
	\bibfield  {author} {\bibinfo {author} {\bibfnamefont {L.}~\bibnamefont
			{Chomaz}},\ }\href@noop {} {\bibinfo {title} {private
			communication}}\BibitemShut {NoStop}%
	\bibitem [{\citenamefont {Li}\ \emph {et~al.}(2017)\citenamefont {Li},
		\citenamefont {Lee}, \citenamefont {Huang}, \citenamefont {Burchesky},
		\citenamefont {Shteynas}, \citenamefont {Top}, \citenamefont {Jamison},\ and\
		\citenamefont {Ketterle}}]{Li2017}%
	\BibitemOpen
	\bibfield  {author} {\bibinfo {author} {\bibfnamefont {J.-R.}\ \bibnamefont
			{Li}}, \bibinfo {author} {\bibfnamefont {J.}~\bibnamefont {Lee}}, \bibinfo
		{author} {\bibfnamefont {W.}~\bibnamefont {Huang}}, \bibinfo {author}
		{\bibfnamefont {S.}~\bibnamefont {Burchesky}}, \bibinfo {author}
		{\bibfnamefont {B.}~\bibnamefont {Shteynas}}, \bibinfo {author}
		{\bibfnamefont {F.~{\c C}.}\ \bibnamefont {Top}}, \bibinfo {author}
		{\bibfnamefont {A.~O.}\ \bibnamefont {Jamison}},\ and\ \bibinfo {author}
		{\bibfnamefont {W.}~\bibnamefont {Ketterle}},\ }\bibfield  {title} {\bibinfo
		{title} {{A stripe phase with supersolid properties in spin--orbit-coupled
				Bose--Einstein condensates}},\ }\href {https://doi.org/10.1038/nature21431}
	{\bibfield  {journal} {\bibinfo  {journal} {Nature}\ }\textbf {\bibinfo
			{volume} {543}},\ \bibinfo {pages} {91} (\bibinfo {year} {2017})}\BibitemShut
	{NoStop}%
	\bibitem [{\citenamefont {Putra}\ \emph {et~al.}(2020)\citenamefont {Putra},
		\citenamefont {Salces-C\'arcoba}, \citenamefont {Yue}, \citenamefont
		{Sugawa},\ and\ \citenamefont {Spielman}}]{Putra2020}%
	\BibitemOpen
	\bibfield  {author} {\bibinfo {author} {\bibfnamefont {A.}~\bibnamefont
			{Putra}}, \bibinfo {author} {\bibfnamefont {F.}~\bibnamefont
			{Salces-C\'arcoba}}, \bibinfo {author} {\bibfnamefont {Y.}~\bibnamefont
			{Yue}}, \bibinfo {author} {\bibfnamefont {S.}~\bibnamefont {Sugawa}},\ and\
		\bibinfo {author} {\bibfnamefont {I.~B.}\ \bibnamefont {Spielman}},\
	}\bibfield  {title} {\bibinfo {title} {{Spatial Coherence of
				Spin-Orbit-Coupled Bose Gases}},\ }\href
	{https://doi.org/10.1103/PhysRevLett.124.053605} {\bibfield  {journal}
		{\bibinfo  {journal} {Phys. Rev. Lett.}\ }\textbf {\bibinfo {volume} {124}},\
		\bibinfo {pages} {053605} (\bibinfo {year} {2020})}\BibitemShut {NoStop}%
	\bibitem [{\citenamefont {Chisholm}\ \emph {et~al.}(2026)\citenamefont
		{Chisholm}, \citenamefont {Hirthe}, \citenamefont {Makhalov}, \citenamefont
		{Ramos}, \citenamefont {Vatré}, \citenamefont {Cabedo}, \citenamefont
		{Celi},\ and\ \citenamefont {Tarruell}}]{Chisholm2026}%
	\BibitemOpen
	\bibfield  {author} {\bibinfo {author} {\bibfnamefont {C.~S.}\ \bibnamefont
			{Chisholm}}, \bibinfo {author} {\bibfnamefont {S.}~\bibnamefont {Hirthe}},
		\bibinfo {author} {\bibfnamefont {V.~B.}\ \bibnamefont {Makhalov}}, \bibinfo
		{author} {\bibfnamefont {R.}~\bibnamefont {Ramos}}, \bibinfo {author}
		{\bibfnamefont {R.}~\bibnamefont {Vatré}}, \bibinfo {author} {\bibfnamefont
			{J.}~\bibnamefont {Cabedo}}, \bibinfo {author} {\bibfnamefont
			{A.}~\bibnamefont {Celi}},\ and\ \bibinfo {author} {\bibfnamefont
			{L.}~\bibnamefont {Tarruell}},\ }\bibfield  {title} {\bibinfo {title}
		{{Probing supersolidity through excitations in a spin-orbit–coupled
				Bose-Einstein condensate}},\ }\href {https://doi.org/10.1126/science.adv1209}
	{\bibfield  {journal} {\bibinfo  {journal} {Science}\ }\textbf {\bibinfo
			{volume} {391}},\ \bibinfo {pages} {480} (\bibinfo {year}
		{2026})}\BibitemShut {NoStop}%
	\bibitem [{\citenamefont {Andreev}\ and\ \citenamefont
		{Lifshitz}(1969)}]{AndreevLifshitz}%
	\BibitemOpen
	\bibfield  {author} {\bibinfo {author} {\bibfnamefont {A.~F.}\ \bibnamefont
			{Andreev}}\ and\ \bibinfo {author} {\bibfnamefont {I.~M.}\ \bibnamefont
			{Lifshitz}},\ }\bibfield  {title} {\bibinfo {title} {Quantum theory of
			defects in crystals},\ }\href
	{http://www.jetp.ras.ru/cgi-bin/e/index/e/29/6/p1107?a=list} {\bibfield
		{journal} {\bibinfo  {journal} {Sov. Phys. JETP}\ }\textbf {\bibinfo {volume}
			{29}},\ \bibinfo {pages} {1107} (\bibinfo {year} {1969})}\BibitemShut
	{NoStop}%
	\bibitem [{\citenamefont {Watanabe}\ and\ \citenamefont
		{Brauner}(2012)}]{Brauner2012}%
	\BibitemOpen
	\bibfield  {author} {\bibinfo {author} {\bibfnamefont {H.}~\bibnamefont
			{Watanabe}}\ and\ \bibinfo {author} {\bibfnamefont {T.}~\bibnamefont
			{Brauner}},\ }\bibfield  {title} {\bibinfo {title} {{Spontaneous breaking of
				continuous translational invariance}},\ }\href
	{https://doi.org/10.1103/PhysRevD.85.085010} {\bibfield  {journal} {\bibinfo
			{journal} {Phys. Rev. D}\ }\textbf {\bibinfo {volume} {85}},\ \bibinfo
		{pages} {085010} (\bibinfo {year} {2012})}\BibitemShut {NoStop}%
	\bibitem [{\citenamefont {Martin}\ \emph {et~al.}(1972)\citenamefont {Martin},
		\citenamefont {Parodi},\ and\ \citenamefont {Pershan}}]{Martin1972}%
	\BibitemOpen
	\bibfield  {author} {\bibinfo {author} {\bibfnamefont {P.~C.}\ \bibnamefont
			{Martin}}, \bibinfo {author} {\bibfnamefont {O.}~\bibnamefont {Parodi}},\
		and\ \bibinfo {author} {\bibfnamefont {P.~S.}\ \bibnamefont {Pershan}},\
	}\bibfield  {title} {\bibinfo {title} {{Unified Hydrodynamic Theory for
				Crystals, Liquid Crystals, and Normal Fluids}},\ }\href
	{https://doi.org/10.1103/PhysRevA.6.2401} {\bibfield  {journal} {\bibinfo
			{journal} {Phys. Rev. A}\ }\textbf {\bibinfo {volume} {6}},\ \bibinfo {pages}
		{2401} (\bibinfo {year} {1972})}\BibitemShut {NoStop}%
	\bibitem [{\citenamefont {Hofmann}\ and\ \citenamefont
		{Zwerger}(2021)}]{Hofmann2021}%
	\BibitemOpen
	\bibfield  {author} {\bibinfo {author} {\bibfnamefont {J.}~\bibnamefont
			{Hofmann}}\ and\ \bibinfo {author} {\bibfnamefont {W.}~\bibnamefont
			{Zwerger}},\ }\bibfield  {title} {\bibinfo {title} {{Hydrodynamics of a
				superfluid smectic}},\ }\href {https://doi.org/10.1088/1742-5468/abe598}
	{\bibfield  {journal} {\bibinfo  {journal} {Journal of Statistical Mechanics:
				Theory and Experiment}\ }\textbf {\bibinfo {volume} {2021}},\ \bibinfo
		{pages} {033104} (\bibinfo {year} {2021})}\BibitemShut {NoStop}%
	\bibitem [{\citenamefont {Zhang}\ \emph
		{et~al.}(2016{\natexlab{a}})\citenamefont {Zhang}, \citenamefont {Yu},
		\citenamefont {Ng}, \citenamefont {Zhang}, \citenamefont {Pitaevskii},\ and\
		\citenamefont {Stringari}}]{Zhang2016}%
	\BibitemOpen
	\bibfield  {author} {\bibinfo {author} {\bibfnamefont {Y.-C.}\ \bibnamefont
			{Zhang}}, \bibinfo {author} {\bibfnamefont {Z.-Q.}\ \bibnamefont {Yu}},
		\bibinfo {author} {\bibfnamefont {T.~K.}\ \bibnamefont {Ng}}, \bibinfo
		{author} {\bibfnamefont {S.}~\bibnamefont {Zhang}}, \bibinfo {author}
		{\bibfnamefont {L.}~\bibnamefont {Pitaevskii}},\ and\ \bibinfo {author}
		{\bibfnamefont {S.}~\bibnamefont {Stringari}},\ }\bibfield  {title} {\bibinfo
		{title} {{Superfluid density of a spin-orbit-coupled Bose gas}},\ }\href
	{https://doi.org/10.1103/PhysRevA.94.033635} {\bibfield  {journal} {\bibinfo
			{journal} {Phys. Rev. A}\ }\textbf {\bibinfo {volume} {94}},\ \bibinfo
		{pages} {033635} (\bibinfo {year} {2016}{\natexlab{a}})}\BibitemShut
	{NoStop}%
	\bibitem [{\citenamefont {Leggett}(1970)}]{Leggett1970}%
	\BibitemOpen
	\bibfield  {author} {\bibinfo {author} {\bibfnamefont {A.~J.}\ \bibnamefont
			{Leggett}},\ }\bibfield  {title} {\bibinfo {title} {{Can a Solid Be
				``Superfluid''?}},\ }\href {https://doi.org/10.1103/PhysRevLett.25.1543}
	{\bibfield  {journal} {\bibinfo  {journal} {Phys. Rev. Lett.}\ }\textbf
		{\bibinfo {volume} {25}},\ \bibinfo {pages} {1543} (\bibinfo {year}
		{1970})}\BibitemShut {NoStop}%
	\bibitem [{\citenamefont {Chauveau}\ \emph {et~al.}(2023)\citenamefont
		{Chauveau}, \citenamefont {Maury}, \citenamefont {Rabec}, \citenamefont
		{Heintze}, \citenamefont {Brochier}, \citenamefont {Nascimbene},
		\citenamefont {Dalibard}, \citenamefont {Beugnon}, \citenamefont {Roccuzzo},\
		and\ \citenamefont {Stringari}}]{Chauveau2023}%
	\BibitemOpen
	\bibfield  {author} {\bibinfo {author} {\bibfnamefont {G.}~\bibnamefont
			{Chauveau}}, \bibinfo {author} {\bibfnamefont {C.}~\bibnamefont {Maury}},
		\bibinfo {author} {\bibfnamefont {F.}~\bibnamefont {Rabec}}, \bibinfo
		{author} {\bibfnamefont {C.}~\bibnamefont {Heintze}}, \bibinfo {author}
		{\bibfnamefont {G.}~\bibnamefont {Brochier}}, \bibinfo {author}
		{\bibfnamefont {S.}~\bibnamefont {Nascimbene}}, \bibinfo {author}
		{\bibfnamefont {J.}~\bibnamefont {Dalibard}}, \bibinfo {author}
		{\bibfnamefont {J.}~\bibnamefont {Beugnon}}, \bibinfo {author} {\bibfnamefont
			{S.~M.}\ \bibnamefont {Roccuzzo}},\ and\ \bibinfo {author} {\bibfnamefont
			{S.}~\bibnamefont {Stringari}},\ }\bibfield  {title} {\bibinfo {title}
		{{Superfluid Fraction in an Interacting Spatially Modulated Bose-Einstein
				Condensate}},\ }\href {https://doi.org/10.1103/PhysRevLett.130.226003}
	{\bibfield  {journal} {\bibinfo  {journal} {Phys. Rev. Lett.}\ }\textbf
		{\bibinfo {volume} {130}},\ \bibinfo {pages} {226003} (\bibinfo {year}
		{2023})}\BibitemShut {NoStop}%
	\bibitem [{\citenamefont {Ripley}\ \emph {et~al.}(2023)\citenamefont {Ripley},
		\citenamefont {Baillie},\ and\ \citenamefont {Blakie}}]{Ripley2023tds}%
	\BibitemOpen
	\bibfield  {author} {\bibinfo {author} {\bibfnamefont {B.~T.~E.}\
			\bibnamefont {Ripley}}, \bibinfo {author} {\bibfnamefont {D.}~\bibnamefont
			{Baillie}},\ and\ \bibinfo {author} {\bibfnamefont {P.~B.}\ \bibnamefont
			{Blakie}},\ }\bibfield  {title} {\bibinfo {title} {{Two-dimensional
				supersolidity in a planar dipolar Bose gas}},\ }\href
	{https://doi.org/10.1103/PhysRevA.108.053321} {\bibfield  {journal} {\bibinfo
			{journal} {Phys. Rev. A}\ }\textbf {\bibinfo {volume} {108}},\ \bibinfo
		{pages} {053321} (\bibinfo {year} {2023})}\BibitemShut {NoStop}%
	\bibitem [{\citenamefont {Tanzi}\ \emph
		{et~al.}(2019{\natexlab{a}})\citenamefont {Tanzi}, \citenamefont {Lucioni},
		\citenamefont {Fam\`a}, \citenamefont {Catani}, \citenamefont {Fioretti},
		\citenamefont {Gabbanini}, \citenamefont {Bisset}, \citenamefont {Santos},\
		and\ \citenamefont {Modugno}}]{F1}%
	\BibitemOpen
	\bibfield  {author} {\bibinfo {author} {\bibfnamefont {L.}~\bibnamefont
			{Tanzi}}, \bibinfo {author} {\bibfnamefont {E.}~\bibnamefont {Lucioni}},
		\bibinfo {author} {\bibfnamefont {F.}~\bibnamefont {Fam\`a}}, \bibinfo
		{author} {\bibfnamefont {J.}~\bibnamefont {Catani}}, \bibinfo {author}
		{\bibfnamefont {A.}~\bibnamefont {Fioretti}}, \bibinfo {author}
		{\bibfnamefont {C.}~\bibnamefont {Gabbanini}}, \bibinfo {author}
		{\bibfnamefont {R.~N.}\ \bibnamefont {Bisset}}, \bibinfo {author}
		{\bibfnamefont {L.}~\bibnamefont {Santos}},\ and\ \bibinfo {author}
		{\bibfnamefont {G.}~\bibnamefont {Modugno}},\ }\bibfield  {title} {\bibinfo
		{title} {{Observation of a Dipolar Quantum Gas with Metastable Supersolid
				Properties}},\ }\href {https://doi.org/10.1103/PhysRevLett.122.130405}
	{\bibfield  {journal} {\bibinfo  {journal} {Phys. Rev. Lett.}\ }\textbf
		{\bibinfo {volume} {122}},\ \bibinfo {pages} {130405} (\bibinfo {year}
		{2019}{\natexlab{a}})}\BibitemShut {NoStop}%
	\bibitem [{\citenamefont {Tanzi}\ \emph
		{et~al.}(2019{\natexlab{b}})\citenamefont {Tanzi}, \citenamefont {Roccuzzo},
		\citenamefont {Lucioni}, \citenamefont {Fam{\`{a}}}, \citenamefont
		{Fioretti}, \citenamefont {Gabbanini}, \citenamefont {Modugno}, \citenamefont
		{Recati},\ and\ \citenamefont {Stringari}}]{F2}%
	\BibitemOpen
	\bibfield  {author} {\bibinfo {author} {\bibfnamefont {L.}~\bibnamefont
			{Tanzi}}, \bibinfo {author} {\bibfnamefont {S.~M.}\ \bibnamefont {Roccuzzo}},
		\bibinfo {author} {\bibfnamefont {E.}~\bibnamefont {Lucioni}}, \bibinfo
		{author} {\bibfnamefont {F.}~\bibnamefont {Fam{\`{a}}}}, \bibinfo {author}
		{\bibfnamefont {A.}~\bibnamefont {Fioretti}}, \bibinfo {author}
		{\bibfnamefont {C.}~\bibnamefont {Gabbanini}}, \bibinfo {author}
		{\bibfnamefont {G.}~\bibnamefont {Modugno}}, \bibinfo {author} {\bibfnamefont
			{A.}~\bibnamefont {Recati}},\ and\ \bibinfo {author} {\bibfnamefont
			{S.}~\bibnamefont {Stringari}},\ }\bibfield  {title} {\bibinfo {title}
		{{Supersolid symmetry breaking from compressional oscillations in a dipolar
				quantum gas}},\ }\href {https://doi.org/10.1038/s41586-019-1568-6} {\bibfield
		{journal} {\bibinfo  {journal} {Nature}\ }\textbf {\bibinfo {volume}
			{574}},\ \bibinfo {pages} {382} (\bibinfo {year}
		{2019}{\natexlab{b}})}\BibitemShut {NoStop}%
	\bibitem [{\citenamefont {B\"ottcher}\ \emph {et~al.}(2019)\citenamefont
		{B\"ottcher}, \citenamefont {Schmidt}, \citenamefont {Wenzel}, \citenamefont
		{Hertkorn}, \citenamefont {Guo}, \citenamefont {Langen},\ and\ \citenamefont
		{Pfau}}]{S1}%
	\BibitemOpen
	\bibfield  {author} {\bibinfo {author} {\bibfnamefont {F.}~\bibnamefont
			{B\"ottcher}}, \bibinfo {author} {\bibfnamefont {J.-N.}\ \bibnamefont
			{Schmidt}}, \bibinfo {author} {\bibfnamefont {M.}~\bibnamefont {Wenzel}},
		\bibinfo {author} {\bibfnamefont {J.}~\bibnamefont {Hertkorn}}, \bibinfo
		{author} {\bibfnamefont {M.}~\bibnamefont {Guo}}, \bibinfo {author}
		{\bibfnamefont {T.}~\bibnamefont {Langen}},\ and\ \bibinfo {author}
		{\bibfnamefont {T.}~\bibnamefont {Pfau}},\ }\bibfield  {title} {\bibinfo
		{title} {{Transient Supersolid Properties in an Array of Dipolar Quantum
				Droplets}},\ }\href {https://doi.org/10.1103/PhysRevX.9.011051} {\bibfield
		{journal} {\bibinfo  {journal} {Phys. Rev. X}\ }\textbf {\bibinfo {volume}
			{9}},\ \bibinfo {pages} {011051} (\bibinfo {year} {2019})}\BibitemShut
	{NoStop}%
	\bibitem [{\citenamefont {Guo}\ \emph {et~al.}(2019)\citenamefont {Guo},
		\citenamefont {B{\"{o}}ttcher}, \citenamefont {Hertkorn}, \citenamefont
		{Schmidt}, \citenamefont {Wenzel}, \citenamefont {B{\"{u}}chler},
		\citenamefont {Langen},\ and\ \citenamefont {Pfau}}]{S2}%
	\BibitemOpen
	\bibfield  {author} {\bibinfo {author} {\bibfnamefont {M.}~\bibnamefont
			{Guo}}, \bibinfo {author} {\bibfnamefont {F.}~\bibnamefont {B{\"{o}}ttcher}},
		\bibinfo {author} {\bibfnamefont {J.}~\bibnamefont {Hertkorn}}, \bibinfo
		{author} {\bibfnamefont {J.-N.}\ \bibnamefont {Schmidt}}, \bibinfo {author}
		{\bibfnamefont {M.}~\bibnamefont {Wenzel}}, \bibinfo {author} {\bibfnamefont
			{H.~P.}\ \bibnamefont {B{\"{u}}chler}}, \bibinfo {author} {\bibfnamefont
			{T.}~\bibnamefont {Langen}},\ and\ \bibinfo {author} {\bibfnamefont
			{T.}~\bibnamefont {Pfau}},\ }\bibfield  {title} {\bibinfo {title} {{The
				low-energy Goldstone mode in a trapped dipolar supersolid}},\ }\href
	{https://doi.org/10.1038/s41586-019-1569-5} {\bibfield  {journal} {\bibinfo
			{journal} {Nature}\ }\textbf {\bibinfo {volume} {574}},\ \bibinfo {pages}
		{386} (\bibinfo {year} {2019})}\BibitemShut {NoStop}%
	\bibitem [{\citenamefont {Chomaz}\ \emph {et~al.}(2019)\citenamefont {Chomaz},
		\citenamefont {Petter}, \citenamefont {Ilzh\"ofer}, \citenamefont {Natale},
		\citenamefont {Trautmann}, \citenamefont {Politi}, \citenamefont
		{Durastante}, \citenamefont {van Bijnen}, \citenamefont {Patscheider},
		\citenamefont {Sohmen}, \citenamefont {Mark},\ and\ \citenamefont
		{Ferlaino}}]{I1}%
	\BibitemOpen
	\bibfield  {author} {\bibinfo {author} {\bibfnamefont {L.}~\bibnamefont
			{Chomaz}}, \bibinfo {author} {\bibfnamefont {D.}~\bibnamefont {Petter}},
		\bibinfo {author} {\bibfnamefont {P.}~\bibnamefont {Ilzh\"ofer}}, \bibinfo
		{author} {\bibfnamefont {G.}~\bibnamefont {Natale}}, \bibinfo {author}
		{\bibfnamefont {A.}~\bibnamefont {Trautmann}}, \bibinfo {author}
		{\bibfnamefont {C.}~\bibnamefont {Politi}}, \bibinfo {author} {\bibfnamefont
			{G.}~\bibnamefont {Durastante}}, \bibinfo {author} {\bibfnamefont {R.~M.~W.}\
			\bibnamefont {van Bijnen}}, \bibinfo {author} {\bibfnamefont
			{A.}~\bibnamefont {Patscheider}}, \bibinfo {author} {\bibfnamefont
			{M.}~\bibnamefont {Sohmen}}, \bibinfo {author} {\bibfnamefont {M.~J.}\
			\bibnamefont {Mark}},\ and\ \bibinfo {author} {\bibfnamefont
			{F.}~\bibnamefont {Ferlaino}},\ }\bibfield  {title} {\bibinfo {title}
		{{Long-Lived and Transient Supersolid Behaviors in Dipolar Quantum Gases}},\
	}\href {https://doi.org/10.1103/PhysRevX.9.021012} {\bibfield  {journal}
		{\bibinfo  {journal} {Phys. Rev. X}\ }\textbf {\bibinfo {volume} {9}},\
		\bibinfo {pages} {021012} (\bibinfo {year} {2019})}\BibitemShut {NoStop}%
	\bibitem [{\citenamefont {Natale}\ \emph {et~al.}(2019)\citenamefont {Natale},
		\citenamefont {van Bijnen}, \citenamefont {Patscheider}, \citenamefont
		{Petter}, \citenamefont {Mark}, \citenamefont {Chomaz},\ and\ \citenamefont
		{Ferlaino}}]{I2}%
	\BibitemOpen
	\bibfield  {author} {\bibinfo {author} {\bibfnamefont {G.}~\bibnamefont
			{Natale}}, \bibinfo {author} {\bibfnamefont {R.~M.~W.}\ \bibnamefont {van
				Bijnen}}, \bibinfo {author} {\bibfnamefont {A.}~\bibnamefont {Patscheider}},
		\bibinfo {author} {\bibfnamefont {D.}~\bibnamefont {Petter}}, \bibinfo
		{author} {\bibfnamefont {M.~J.}\ \bibnamefont {Mark}}, \bibinfo {author}
		{\bibfnamefont {L.}~\bibnamefont {Chomaz}},\ and\ \bibinfo {author}
		{\bibfnamefont {F.}~\bibnamefont {Ferlaino}},\ }\bibfield  {title} {\bibinfo
		{title} {{Excitation Spectrum of a Trapped Dipolar Supersolid and Its
				Experimental Evidence}},\ }\href
	{https://doi.org/10.1103/PhysRevLett.123.050402} {\bibfield  {journal}
		{\bibinfo  {journal} {Phys. Rev. Lett.}\ }\textbf {\bibinfo {volume} {123}},\
		\bibinfo {pages} {050402} (\bibinfo {year} {2019})}\BibitemShut {NoStop}%
	\bibitem [{\citenamefont {Norcia}\ \emph {et~al.}(2021)\citenamefont {Norcia},
		\citenamefont {Politi}, \citenamefont {Klaus}, \citenamefont {Poli},
		\citenamefont {Sohmen}, \citenamefont {Mark}, \citenamefont {Bisset},
		\citenamefont {Santos},\ and\ \citenamefont {Ferlaino}}]{Norcia2021_2D}%
	\BibitemOpen
	\bibfield  {author} {\bibinfo {author} {\bibfnamefont {M.~A.}\ \bibnamefont
			{Norcia}}, \bibinfo {author} {\bibfnamefont {C.}~\bibnamefont {Politi}},
		\bibinfo {author} {\bibfnamefont {L.}~\bibnamefont {Klaus}}, \bibinfo
		{author} {\bibfnamefont {E.}~\bibnamefont {Poli}}, \bibinfo {author}
		{\bibfnamefont {M.}~\bibnamefont {Sohmen}}, \bibinfo {author} {\bibfnamefont
			{M.~J.}\ \bibnamefont {Mark}}, \bibinfo {author} {\bibfnamefont {R.~N.}\
			\bibnamefont {Bisset}}, \bibinfo {author} {\bibfnamefont {L.}~\bibnamefont
			{Santos}},\ and\ \bibinfo {author} {\bibfnamefont {F.}~\bibnamefont
			{Ferlaino}},\ }\bibfield  {title} {\bibinfo {title} {Two-dimensional
			supersolidity in a dipolar quantum gas},\ }\href
	{https://doi.org/10.1038/s41586-021-03725-7} {\bibfield  {journal} {\bibinfo
			{journal} {Nature}\ }\textbf {\bibinfo {volume} {596}},\ \bibinfo {pages}
		{357} (\bibinfo {year} {2021})}\BibitemShut {NoStop}%
	\bibitem [{\citenamefont {Bland}\ \emph {et~al.}(2022)\citenamefont {Bland},
		\citenamefont {Poli}, \citenamefont {Politi}, \citenamefont {Klaus},
		\citenamefont {Norcia}, \citenamefont {Ferlaino}, \citenamefont {Santos},\
		and\ \citenamefont {Bisset}}]{Bland2021_2D}%
	\BibitemOpen
	\bibfield  {author} {\bibinfo {author} {\bibfnamefont {T.}~\bibnamefont
			{Bland}}, \bibinfo {author} {\bibfnamefont {E.}~\bibnamefont {Poli}},
		\bibinfo {author} {\bibfnamefont {C.}~\bibnamefont {Politi}}, \bibinfo
		{author} {\bibfnamefont {L.}~\bibnamefont {Klaus}}, \bibinfo {author}
		{\bibfnamefont {M.~A.}\ \bibnamefont {Norcia}}, \bibinfo {author}
		{\bibfnamefont {F.}~\bibnamefont {Ferlaino}}, \bibinfo {author}
		{\bibfnamefont {L.}~\bibnamefont {Santos}},\ and\ \bibinfo {author}
		{\bibfnamefont {R.~N.}\ \bibnamefont {Bisset}},\ }\bibfield  {title}
	{\bibinfo {title} {{Two-Dimensional Supersolid Formation in Dipolar
				Condensates}},\ }\href {https://doi.org/10.1103/PhysRevLett.128.195302}
	{\bibfield  {journal} {\bibinfo  {journal} {Phys. Rev. Lett.}\ }\textbf
		{\bibinfo {volume} {128}},\ \bibinfo {pages} {195302} (\bibinfo {year}
		{2022})}\BibitemShut {NoStop}%
	\bibitem [{\citenamefont {Lima}\ \emph {et~al.}(2025)\citenamefont {Lima},
		\citenamefont {Grossklags}, \citenamefont {Zampronio}, \citenamefont
		{Cinti},\ and\ \citenamefont {Mendoza-Coto}}]{lima2025sdp}%
	\BibitemOpen
	\bibfield  {author} {\bibinfo {author} {\bibfnamefont {D.}~\bibnamefont
			{Lima}}, \bibinfo {author} {\bibfnamefont {M.}~\bibnamefont {Grossklags}},
		\bibinfo {author} {\bibfnamefont {V.}~\bibnamefont {Zampronio}}, \bibinfo
		{author} {\bibfnamefont {F.}~\bibnamefont {Cinti}},\ and\ \bibinfo {author}
		{\bibfnamefont {A.}~\bibnamefont {Mendoza-Coto}},\ }\bibfield  {title}
	{\bibinfo {title} {{Supersolid dipolar phases in planar geometry: Effects of
				tilted polarization}},\ }\href {https://doi.org/10.1103/mlfn-114m} {\bibfield
		{journal} {\bibinfo  {journal} {Phys. Rev. A}\ }\textbf {\bibinfo {volume}
			{111}},\ \bibinfo {pages} {063311} (\bibinfo {year} {2025})}\BibitemShut
	{NoStop}%
	\bibitem [{\citenamefont {Yoo}\ and\ \citenamefont
		{Dorsey}(2010)}]{Dorsey2010}%
	\BibitemOpen
	\bibfield  {author} {\bibinfo {author} {\bibfnamefont {C.-D.}\ \bibnamefont
			{Yoo}}\ and\ \bibinfo {author} {\bibfnamefont {A.~T.}\ \bibnamefont
			{Dorsey}},\ }\bibfield  {title} {\bibinfo {title} {{Hydrodynamic theory of
				supersolids: Variational principle, effective Lagrangian, and density-density
				correlation function}},\ }\href {https://doi.org/10.1103/PhysRevB.81.134518}
	{\bibfield  {journal} {\bibinfo  {journal} {Phys. Rev. B}\ }\textbf {\bibinfo
			{volume} {81}},\ \bibinfo {pages} {134518} (\bibinfo {year}
		{2010})}\BibitemShut {NoStop}%
	\bibitem [{\citenamefont {Zawi\ifmmode~\acute{s}\else \'{s}\fi{}lak}\ \emph
		{et~al.}(2025)\citenamefont {Zawi\ifmmode~\acute{s}\else \'{s}\fi{}lak},
		\citenamefont {\ifmmode~\check{S}\else \v{S}\fi{}indik}, \citenamefont
		{Stringari},\ and\ \citenamefont {Recati}}]{ZawiAnomalous}%
	\BibitemOpen
	\bibfield  {author} {\bibinfo {author} {\bibfnamefont {T.}~\bibnamefont
			{Zawi\ifmmode~\acute{s}\else \'{s}\fi{}lak}}, \bibinfo {author}
		{\bibfnamefont {M.}~\bibnamefont {\ifmmode~\check{S}\else \v{S}\fi{}indik}},
		\bibinfo {author} {\bibfnamefont {S.}~\bibnamefont {Stringari}},\ and\
		\bibinfo {author} {\bibfnamefont {A.}~\bibnamefont {Recati}},\ }\bibfield
	{title} {\bibinfo {title} {{Anomalous Doppler Effect in Superfluid and
				Supersolid Atomic Gases}},\ }\href
	{https://doi.org/10.1103/PhysRevLett.134.226001} {\bibfield  {journal}
		{\bibinfo  {journal} {Phys. Rev. Lett.}\ }\textbf {\bibinfo {volume} {134}},\
		\bibinfo {pages} {226001} (\bibinfo {year} {2025})}\BibitemShut {NoStop}%
	\bibitem [{\citenamefont {De~Gennes}\ and\ \citenamefont
		{Prost}(1993)}]{DeGennesBookLC}%
	\BibitemOpen
	\bibfield  {author} {\bibinfo {author} {\bibfnamefont {P.~G.}\ \bibnamefont
			{De~Gennes}}\ and\ \bibinfo {author} {\bibfnamefont {J.}~\bibnamefont
			{Prost}},\ }\href@noop {} { {\bibinfo {title} {{The Physics of Liquid
					Crystals}}}}\ (\bibinfo  {publisher} {CLARENDON PRESS - OXFORD},\ \bibinfo
	{year} {1993})\BibitemShut {NoStop}%
	\bibitem [{\citenamefont {Blakie}\ \emph {et~al.}(2023)\citenamefont {Blakie},
		\citenamefont {Chomaz}, \citenamefont {Baillie},\ and\ \citenamefont
		{Ferlaino}}]{Blakie2023sounds}%
	\BibitemOpen
	\bibfield  {author} {\bibinfo {author} {\bibfnamefont {P.~B.}\ \bibnamefont
			{Blakie}}, \bibinfo {author} {\bibfnamefont {L.}~\bibnamefont {Chomaz}},
		\bibinfo {author} {\bibfnamefont {D.}~\bibnamefont {Baillie}},\ and\ \bibinfo
		{author} {\bibfnamefont {F.}~\bibnamefont {Ferlaino}},\ }\bibfield  {title}
	{\bibinfo {title} {{Compressibility and speeds of sound across the
				superfluid-to-supersolid phase transition of an elongated dipolar gas}},\
	}\href {https://doi.org/10.1103/PhysRevResearch.5.033161} {\bibfield
		{journal} {\bibinfo  {journal} {Phys. Rev. Res.}\ }\textbf {\bibinfo {volume}
			{5}},\ \bibinfo {pages} {033161} (\bibinfo {year} {2023})}\BibitemShut
	{NoStop}%
	\bibitem [{\citenamefont {W\"achtler}\ and\ \citenamefont
		{Santos}(2016)}]{Wachter2016}%
	\BibitemOpen
	\bibfield  {author} {\bibinfo {author} {\bibfnamefont {F.}~\bibnamefont
			{W\"achtler}}\ and\ \bibinfo {author} {\bibfnamefont {L.}~\bibnamefont
			{Santos}},\ }\bibfield  {title} {\bibinfo {title} {{Quantum filaments in
				dipolar Bose-Einstein condensates}},\ }\href
	{https://doi.org/10.1103/PhysRevA.93.061603} {\bibfield  {journal} {\bibinfo
			{journal} {Phys. Rev. A}\ }\textbf {\bibinfo {volume} {93}},\ \bibinfo
		{pages} {061603} (\bibinfo {year} {2016})}\BibitemShut {NoStop}%
	\bibitem [{\citenamefont {\ifmmode~\check{S}\else \v{S}\fi{}indik}\ \emph
		{et~al.}(2024)\citenamefont {\ifmmode~\check{S}\else \v{S}\fi{}indik},
		\citenamefont {Zawi\ifmmode~\acute{s}\else \'{s}\fi{}lak}, \citenamefont
		{Recati},\ and\ \citenamefont {Stringari}}]{Sindik2024}%
	\BibitemOpen
	\bibfield  {author} {\bibinfo {author} {\bibfnamefont {M.}~\bibnamefont
			{\ifmmode~\check{S}\else \v{S}\fi{}indik}}, \bibinfo {author} {\bibfnamefont
			{T.}~\bibnamefont {Zawi\ifmmode~\acute{s}\else \'{s}\fi{}lak}}, \bibinfo
		{author} {\bibfnamefont {A.}~\bibnamefont {Recati}},\ and\ \bibinfo {author}
		{\bibfnamefont {S.}~\bibnamefont {Stringari}},\ }\bibfield  {title} {\bibinfo
		{title} {{Sound, Superfluidity, and Layer Compressibility in a Ring Dipolar
				Supersolid}},\ }\href {https://doi.org/10.1103/PhysRevLett.132.146001}
	{\bibfield  {journal} {\bibinfo  {journal} {Phys. Rev. Lett.}\ }\textbf
		{\bibinfo {volume} {132}},\ \bibinfo {pages} {146001} (\bibinfo {year}
		{2024})}\BibitemShut {NoStop}%
	\bibitem [{\citenamefont {Platt}\ \emph {et~al.}(2024)\citenamefont {Platt},
		\citenamefont {Baillie},\ and\ \citenamefont {Blakie}}]{Platt2024}%
	\BibitemOpen
	\bibfield  {author} {\bibinfo {author} {\bibfnamefont {L.~M.}\ \bibnamefont
			{Platt}}, \bibinfo {author} {\bibfnamefont {D.}~\bibnamefont {Baillie}},\
		and\ \bibinfo {author} {\bibfnamefont {P.~B.}\ \bibnamefont {Blakie}},\
	}\bibfield  {title} {\bibinfo {title} {{Sound waves and fluctuations in
				one-dimensional supersolids}},\ }\href
	{https://doi.org/10.1103/PhysRevA.110.023320} {\bibfield  {journal} {\bibinfo
			{journal} {Phys. Rev. A}\ }\textbf {\bibinfo {volume} {110}},\ \bibinfo
		{pages} {023320} (\bibinfo {year} {2024})}\BibitemShut {NoStop}%
	\bibitem [{Note1()}]{Note1}%
	\BibitemOpen
	\bibinfo {note} {Notice that the speed of the transverse sound in an
		isotropic (super)solid has a similar expression, where $\lambda _y$ is
		replaced by the second Lamé coefficient~\cite
		{Dorsey2010,Poli2024eoa}.}\BibitemShut {Stop}%
	\bibitem [{\citenamefont {Lin}\ \emph {et~al.}(2011)\citenamefont {Lin},
		\citenamefont {Jiménez-García},\ and\ \citenamefont {Spielman}}]{Lin2011}%
	\BibitemOpen
	\bibfield  {author} {\bibinfo {author} {\bibfnamefont {Y.-J.}\ \bibnamefont
			{Lin}}, \bibinfo {author} {\bibfnamefont {K.}~\bibnamefont
			{Jiménez-García}},\ and\ \bibinfo {author} {\bibfnamefont {I.~B.}\
			\bibnamefont {Spielman}},\ }\bibfield  {title} {\bibinfo {title}
		{{Spin-orbit-coupled Bose-Einstein condensates}},\ }\href
	{https://doi.org/10.1038/nature09887} {\bibfield  {journal} {\bibinfo
			{journal} {Nature}\ }\textbf {\bibinfo {volume} {471}},\ \bibinfo {pages}
		{83} (\bibinfo {year} {2011})}\BibitemShut {NoStop}%
	\bibitem [{\citenamefont {Pitaevskii}\ and\ \citenamefont
		{Stringari}(2016)}]{BecBook2016}%
	\BibitemOpen
	\bibfield  {author} {\bibinfo {author} {\bibfnamefont {L.}~\bibnamefont
			{Pitaevskii}}\ and\ \bibinfo {author} {\bibfnamefont {S.}~\bibnamefont
			{Stringari}},\ }\href@noop {} { {\bibinfo {title} {{Bose-Einstein
					condensation and superfluidity}}}}\ (\bibinfo  {publisher} {Oxford University
		Press},\ \bibinfo {year} {2016})\BibitemShut {NoStop}%
	\bibitem [{\citenamefont {Zhai}(2015)}]{Zhai2015review}%
	\BibitemOpen
	\bibfield  {author} {\bibinfo {author} {\bibfnamefont {H.}~\bibnamefont
			{Zhai}},\ }\bibfield  {title} {\bibinfo {title} {{Degenerate quantum gases
				with spin–orbit coupling: a review}},\ }\href
	{https://doi.org/10.1088/0034-4885/78/2/026001} {\bibfield  {journal}
		{\bibinfo  {journal} {Reports on Progress in Physics}\ }\textbf {\bibinfo
			{volume} {78}},\ \bibinfo {pages} {026001} (\bibinfo {year}
		{2015})}\BibitemShut {NoStop}%
	\bibitem [{\citenamefont {Zhang}\ \emph
		{et~al.}(2016{\natexlab{b}})\citenamefont {Zhang}, \citenamefont {Mossman},
		\citenamefont {Busch}, \citenamefont {Engels},\ and\ \citenamefont
		{Zhang}}]{Zhang2016review}%
	\BibitemOpen
	\bibfield  {author} {\bibinfo {author} {\bibfnamefont {Y.}~\bibnamefont
			{Zhang}}, \bibinfo {author} {\bibfnamefont {M.~E.}\ \bibnamefont {Mossman}},
		\bibinfo {author} {\bibfnamefont {T.}~\bibnamefont {Busch}}, \bibinfo
		{author} {\bibfnamefont {P.}~\bibnamefont {Engels}},\ and\ \bibinfo {author}
		{\bibfnamefont {C.}~\bibnamefont {Zhang}},\ }\bibfield  {title} {\bibinfo
		{title} {{Properties of spin--orbit-coupled Bose--Einstein condensates}},\
	}\href {https://doi.org/10.1007/s11467-016-0560-y} {\bibfield  {journal}
		{\bibinfo  {journal} {Frontiers of Physics}\ }\textbf {\bibinfo {volume}
			{11}},\ \bibinfo {pages} {118103} (\bibinfo {year}
		{2016}{\natexlab{b}})}\BibitemShut {NoStop}%
	\bibitem [{\citenamefont {Martone}\ and\ \citenamefont
		{Stringari}(2025)}]{Martone2025review}%
	\BibitemOpen
	\bibfield  {author} {\bibinfo {author} {\bibfnamefont {G.~I.}\ \bibnamefont
			{Martone}}\ and\ \bibinfo {author} {\bibfnamefont {S.}~\bibnamefont
			{Stringari}},\ }\bibfield  {title} {\bibinfo {title} {{Two-component
				ultracold Bose gases with spin-orbit coupling}},\ }in\ \href@noop {} {
		{\bibinfo {booktitle} {Quantum Mixtures with Ultra-cold Atoms}}}\ (\bibinfo
	{publisher} {IOS Press},\ \bibinfo {year} {2025})\ pp.\ \bibinfo {pages}
	{53--73}\BibitemShut {NoStop}%
	\bibitem [{\citenamefont {Martone}\ and\ \citenamefont
		{Shlyapnikov}(2026)}]{Martone2026a}%
	\BibitemOpen
	\bibfield  {author} {\bibinfo {author} {\bibfnamefont {G.~I.}\ \bibnamefont
			{Martone}}\ and\ \bibinfo {author} {\bibfnamefont {G.~V.}\ \bibnamefont
			{Shlyapnikov}},\ }\bibfield  {title} {\bibinfo {title} {{Traveling supersolid
				stripe patterns in spin-orbit-coupled Bose-Einstein condensates}},\ }\href
	{https://doi.org/10.1103/wf3t-h862} {\bibfield  {journal} {\bibinfo
			{journal} {Phys. Rev. A}\ }\textbf {\bibinfo {volume} {113}},\ \bibinfo
		{pages} {013331} (\bibinfo {year} {2026})}\BibitemShut {NoStop}%
	\bibitem [{Note2()}]{Note2}%
	\BibitemOpen
	\bibinfo {note} {In general, the superflow velocity is given by the deviation
		of $\protect \mathbf {v}_\protect \mathrm {s}$ from its ground-state value
		$\hbar (k_+ + k_-) \protect \hat {\protect \mathbf {x}} / 2 m$. In the
		$\protect \mathbb {Z}_2$ case studied in the following the ground state value
		is zero.}\BibitemShut {Stop}%
	\bibitem [{Note3()}]{Note3}%
	\BibitemOpen
	\bibinfo {note} {Although all the quantities have been introduced using a
		two-harmonic approximation for $\Psi $, generalizing their definitions to
		include higher-order terms induced by the interactions is
		straightforward~\cite {Martone2026a,Martone2026b}.}\BibitemShut {Stop}%
	\bibitem [{\citenamefont {Martone}()}]{Martone2026b}%
	\BibitemOpen
	\bibfield  {author} {\bibinfo {author} {\bibfnamefont {G.~I.}\ \bibnamefont
			{Martone}},\ }\href@noop {} {\bibinfo {title} {{Hydrodynamic theory of
				supersolid spin-orbit-coupled Bose gases}}},\ \bibinfo {note} {in
		preparation}\BibitemShut {NoStop}%
	\bibitem [{\citenamefont {Baym}(1968)}]{baym1968}%
	\BibitemOpen
	\bibfield  {author} {\bibinfo {author} {\bibfnamefont {G.}~\bibnamefont
		{Baym}},\ }\bibinfo {title} {The microscopic description of superfluidity},\
		in\ { {\bibinfo
		{booktitle} {Mathematical Methods in Solid State and\ Superfluid\ Theory}}},\
	\bibinfo {editor} {edited by\ \bibinfo {editor} {\bibfnamefont {R.~C.}\
		\bibnamefont {Clark}}\ and\ \bibinfo {editor} {\bibfnamefont {G.~H.}\
		\bibnamefont {Derrick}}}\ (\bibinfo  {publisher} {Springer US},\ \bibinfo
	{address} {Boston, MA},\ \bibinfo {year} {1968})\ pp.\ \bibinfo {pages}
	{121--156}\BibitemShut {NoStop}%
	\bibitem [{\citenamefont {Scalapino}\ \emph {et~al.}(1992)\citenamefont
		{Scalapino}, \citenamefont {White},\ and\ \citenamefont
		{Zhang}}]{scalapino92}%
	\BibitemOpen
	\bibfield  {author} {\bibinfo {author} {\bibfnamefont {D.~J.}\ \bibnamefont
			{Scalapino}}, \bibinfo {author} {\bibfnamefont {S.~R.}\ \bibnamefont
			{White}},\ and\ \bibinfo {author} {\bibfnamefont {S.~C.}\ \bibnamefont
			{Zhang}},\ }\bibfield  {title} {\bibinfo {title} {{Superfluid density and the
				Drude weight of the Hubbard model}},\ }\href
	{https://doi.org/10.1103/PhysRevLett.68.2830} {\bibfield  {journal} {\bibinfo
			{journal} {Phys. Rev. Lett.}\ }\textbf {\bibinfo {volume} {68}},\ \bibinfo
		{pages} {2830} (\bibinfo {year} {1992})}\BibitemShut {NoStop}%
	\bibitem [{\citenamefont {Saslow}(2025)}]{Saslow2025}%
	\BibitemOpen
	\bibfield  {author} {\bibinfo {author} {\bibfnamefont {W.~M.}\ \bibnamefont
			{Saslow}},\ }\bibfield  {title} {\bibinfo {title} {{Dynamics of supersolid
				state: Normal fluid, superfluid, and supersolid velocities}},\ }\href
	{https://doi.org/10.1103/yyjt-1tth} {\bibfield  {journal} {\bibinfo
			{journal} {Phys. Rev. A}\ }\textbf {\bibinfo {volume} {112}},\ \bibinfo
		{pages} {033303} (\bibinfo {year} {2025})}\BibitemShut {NoStop}%
	\bibitem [{\citenamefont {Martone}\ \emph {et~al.}(2012)\citenamefont
		{Martone}, \citenamefont {Li}, \citenamefont {Pitaevskii},\ and\
		\citenamefont {Stringari}}]{Martone2012}%
	\BibitemOpen
	\bibfield  {author} {\bibinfo {author} {\bibfnamefont {G.~I.}\ \bibnamefont
			{Martone}}, \bibinfo {author} {\bibfnamefont {Y.}~\bibnamefont {Li}},
		\bibinfo {author} {\bibfnamefont {L.~P.}\ \bibnamefont {Pitaevskii}},\ and\
		\bibinfo {author} {\bibfnamefont {S.}~\bibnamefont {Stringari}},\ }\bibfield
	{title} {\bibinfo {title} {{Anisotropic dynamics of a spin-orbit-coupled
				Bose-Einstein condensate}},\ }\href
	{https://doi.org/10.1103/PhysRevA.86.063621} {\bibfield  {journal} {\bibinfo
			{journal} {Phys. Rev. A}\ }\textbf {\bibinfo {volume} {86}},\ \bibinfo
		{pages} {063621} (\bibinfo {year} {2012})}\BibitemShut {NoStop}%
	\bibitem [{\citenamefont {Li}\ \emph {et~al.}(2013)\citenamefont {Li},
		\citenamefont {Martone}, \citenamefont {Pitaevskii},\ and\ \citenamefont
		{Stringari}}]{Li2013}%
	\BibitemOpen
	\bibfield  {author} {\bibinfo {author} {\bibfnamefont {Y.}~\bibnamefont
			{Li}}, \bibinfo {author} {\bibfnamefont {G.~I.}\ \bibnamefont {Martone}},
		\bibinfo {author} {\bibfnamefont {L.~P.}\ \bibnamefont {Pitaevskii}},\ and\
		\bibinfo {author} {\bibfnamefont {S.}~\bibnamefont {Stringari}},\ }\bibfield
	{title} {\bibinfo {title} {{Superstripes and the Excitation Spectrum of a
				Spin-Orbit-Coupled Bose-Einstein Condensate}},\ }\href
	{https://doi.org/10.1103/PhysRevLett.110.235302} {\bibfield  {journal}
		{\bibinfo  {journal} {Phys. Rev. Lett.}\ }\textbf {\bibinfo {volume} {110}},\
		\bibinfo {pages} {235302} (\bibinfo {year} {2013})}\BibitemShut {NoStop}%
	\bibitem [{\citenamefont {Jian}\ and\ \citenamefont {Zhai}(2011)}]{Jian2011}%
	\BibitemOpen
	\bibfield  {author} {\bibinfo {author} {\bibfnamefont {C.-M.}\ \bibnamefont
			{Jian}}\ and\ \bibinfo {author} {\bibfnamefont {H.}~\bibnamefont {Zhai}},\
	}\bibfield  {title} {\bibinfo {title} {{Paired superfluidity and
				fractionalized vortices in systems of spin-orbit coupled bosons}},\ }\href
	{https://doi.org/10.1103/PhysRevB.84.060508} {\bibfield  {journal} {\bibinfo
			{journal} {Phys. Rev. B}\ }\textbf {\bibinfo {volume} {84}},\ \bibinfo
		{pages} {060508} (\bibinfo {year} {2011})}\BibitemShut {NoStop}%
	\bibitem [{\citenamefont {Geier}\ \emph {et~al.}(2023)\citenamefont {Geier},
		\citenamefont {Martone}, \citenamefont {Hauke}, \citenamefont {Ketterle},\
		and\ \citenamefont {Stringari}}]{Geier2023}%
	\BibitemOpen
	\bibfield  {author} {\bibinfo {author} {\bibfnamefont {K.~T.}\ \bibnamefont
			{Geier}}, \bibinfo {author} {\bibfnamefont {G.~I.}\ \bibnamefont {Martone}},
		\bibinfo {author} {\bibfnamefont {P.}~\bibnamefont {Hauke}}, \bibinfo
		{author} {\bibfnamefont {W.}~\bibnamefont {Ketterle}},\ and\ \bibinfo
		{author} {\bibfnamefont {S.}~\bibnamefont {Stringari}},\ }\bibfield  {title}
	{\bibinfo {title} {{Dynamics of Stripe Patterns in Supersolid
				Spin-Orbit-Coupled Bose Gases}},\ }\href
	{https://doi.org/10.1103/PhysRevLett.130.156001} {\bibfield  {journal}
		{\bibinfo  {journal} {Phys. Rev. Lett.}\ }\textbf {\bibinfo {volume} {130}},\
		\bibinfo {pages} {156001} (\bibinfo {year} {2023})}\BibitemShut {NoStop}%
	\bibitem [{\citenamefont {Hegg}\ \emph {et~al.}(2024)\citenamefont {Hegg},
		\citenamefont {Jiang}, \citenamefont {Wang}, \citenamefont {Hou},
		\citenamefont {Zeng}, \citenamefont {Yildirim},\ and\ \citenamefont
		{Ku}}]{hegg2024}%
	\BibitemOpen
	\bibfield  {author} {\bibinfo {author} {\bibfnamefont {A.}~\bibnamefont
			{Hegg}}, \bibinfo {author} {\bibfnamefont {R.}~\bibnamefont {Jiang}},
		\bibinfo {author} {\bibfnamefont {J.}~\bibnamefont {Wang}}, \bibinfo {author}
		{\bibfnamefont {J.}~\bibnamefont {Hou}}, \bibinfo {author} {\bibfnamefont
			{T.}~\bibnamefont {Zeng}}, \bibinfo {author} {\bibfnamefont {Y.}~\bibnamefont
			{Yildirim}},\ and\ \bibinfo {author} {\bibfnamefont {W.}~\bibnamefont {Ku}},\
	}\bibfield  {title} {\bibinfo {title} {{Universal low-temperature fluctuation
				of unconventional superconductors revealed: ``Smoking gun'' leaves proper
				bosonic superfluidity the last theory standing}},\ }\href
	{https://arxiv.org/abs/2402.08730} {\bibfield  {journal} {\bibinfo  {journal}
			{arXiv:2402.08730 [cond-mat.supr-con]}\ } (\bibinfo {year}
		{2024})}\BibitemShut {NoStop}%
	\bibitem [{\citenamefont {Berger}\ \emph {et~al.}(2025)\citenamefont {Berger},
		\citenamefont {Prokof'ev},\ and\ \citenamefont {Svistunov}}]{berger2025}%
	\BibitemOpen
	\bibfield  {author} {\bibinfo {author} {\bibfnamefont {V.}~\bibnamefont
			{Berger}}, \bibinfo {author} {\bibfnamefont {N.}~\bibnamefont {Prokof'ev}},\
		and\ \bibinfo {author} {\bibfnamefont {B.}~\bibnamefont {Svistunov}},\
	}\bibfield  {title} {\bibinfo {title} {{``Depletion''' of Superfluid Density:
				Universal Low-temperature Thermodynamics of Superfluids}},\ }\href
	{https://arxiv.org/abs/2506.22683} {\bibfield  {journal} {\bibinfo  {journal}
			{arXiv:2506.22683 [cond-mat.quant-gas]}\ } (\bibinfo {year}
		{2025})}\BibitemShut {NoStop}%
	\bibitem [{\citenamefont {Cook}\ \emph {et~al.}(2026)\citenamefont {Cook},
		\citenamefont {Lee},\ and\ \citenamefont {Blakie}}]{Cook2026}%
	\BibitemOpen
	\bibfield  {author} {\bibinfo {author} {\bibfnamefont {R.}~\bibnamefont
			{Cook}}, \bibinfo {author} {\bibfnamefont {A.-C.}\ \bibnamefont {Lee}},\ and\
		\bibinfo {author} {\bibfnamefont {P.~B.}\ \bibnamefont {Blakie}},\ }\bibfield
	{title} {\bibinfo {title} {{Excitations and anisotropic sound in planar
				dipolar supersolids with tilted dipoles}},\ }\href
	{https://arxiv.org/abs/2602.01617} {\bibfield  {journal} {\bibinfo  {journal}
			{arXiv:2602.01617 [cond-mat.quant-gas]}\ } (\bibinfo {year}
		{2026})}\BibitemShut {NoStop}%
	\bibitem [{\citenamefont {Poli}\ \emph {et~al.}(2024)\citenamefont {Poli},
		\citenamefont {Baillie}, \citenamefont {Ferlaino},\ and\ \citenamefont
		{Blakie}}]{Poli2024eoa}%
	\BibitemOpen
	\bibfield  {author} {\bibinfo {author} {\bibfnamefont {E.}~\bibnamefont
			{Poli}}, \bibinfo {author} {\bibfnamefont {D.}~\bibnamefont {Baillie}},
		\bibinfo {author} {\bibfnamefont {F.}~\bibnamefont {Ferlaino}},\ and\
		\bibinfo {author} {\bibfnamefont {P.~B.}\ \bibnamefont {Blakie}},\ }\bibfield
	{title} {\bibinfo {title} {{Excitations of a two-dimensional supersolid}},\
	}\href {https://doi.org/10.1103/PhysRevA.110.053301} {\bibfield  {journal}
		{\bibinfo  {journal} {Phys. Rev. A}\ }\textbf {\bibinfo {volume} {110}},\
		\bibinfo {pages} {053301} (\bibinfo {year} {2024})}\BibitemShut {NoStop}%
	\bibitem [{\citenamefont {Sch\"utzhold}\ \emph {et~al.}(2006)\citenamefont
		{Sch\"utzhold}, \citenamefont {Uhlmann}, \citenamefont {Xu},\ and\
		\citenamefont {Fischer}}]{FischerLHY}%
	\BibitemOpen
	\bibfield  {author} {\bibinfo {author} {\bibfnamefont {R.}~\bibnamefont
			{Sch\"utzhold}}, \bibinfo {author} {\bibfnamefont {M.}~\bibnamefont
			{Uhlmann}}, \bibinfo {author} {\bibfnamefont {Y.}~\bibnamefont {Xu}},\ and\
		\bibinfo {author} {\bibfnamefont {U.~R.}\ \bibnamefont {Fischer}},\
	}\bibfield  {title} {\bibinfo {title} {{Mean-field expansion in Bose-Einstein
				condensates with finite range interactions}},\ }\href
	{https://doi.org/10.1142/S0217979206035631} {\bibfield  {journal} {\bibinfo
			{journal} {International Journal of Modern Physics B}\ }\textbf {\bibinfo
			{volume} {20}},\ \bibinfo {pages} {3555} (\bibinfo {year}
		{2006})}\BibitemShut {NoStop}%
	\bibitem [{\citenamefont {Lima}\ and\ \citenamefont {Pelster}(2012)}]{Pelster}%
	\BibitemOpen
	\bibfield  {author} {\bibinfo {author} {\bibfnamefont {A.~R.~P.}\
			\bibnamefont {Lima}}\ and\ \bibinfo {author} {\bibfnamefont {A.}~\bibnamefont
			{Pelster}},\ }\bibfield  {title} {\bibinfo {title} {{Beyond mean-field
				low-lying excitations of dipolar Bose gases}},\ }\href
	{https://doi.org/10.1103/PhysRevA.86.063609} {\bibfield  {journal} {\bibinfo
			{journal} {Phys. Rev. A}\ }\textbf {\bibinfo {volume} {86}},\ \bibinfo
		{pages} {063609} (\bibinfo {year} {2012})}\BibitemShut {NoStop}%
	\bibitem [{\citenamefont {Bisset}\ \emph {et~al.}(2016)\citenamefont {Bisset},
		\citenamefont {Wilson}, \citenamefont {Baillie},\ and\ \citenamefont
		{Blakie}}]{Bisset2016gsp}%
	\BibitemOpen
	\bibfield  {author} {\bibinfo {author} {\bibfnamefont {R.~N.}\ \bibnamefont
			{Bisset}}, \bibinfo {author} {\bibfnamefont {R.~M.}\ \bibnamefont {Wilson}},
		\bibinfo {author} {\bibfnamefont {D.}~\bibnamefont {Baillie}},\ and\ \bibinfo
		{author} {\bibfnamefont {P.~B.}\ \bibnamefont {Blakie}},\ }\bibfield  {title}
	{\bibinfo {title} {{Ground-state phase diagram of a dipolar condensate with
				quantum fluctuations}},\ }\href {https://doi.org/10.1103/PhysRevA.94.033619}
	{\bibfield  {journal} {\bibinfo  {journal} {Phys. Rev. A}\ }\textbf {\bibinfo
			{volume} {94}},\ \bibinfo {pages} {033619} (\bibinfo {year}
		{2016})}\BibitemShut {NoStop}%
	\bibitem [{\citenamefont {Martone}\ and\ \citenamefont
		{Stringari}(2021)}]{Martone2021}%
	\BibitemOpen
	\bibfield  {author} {\bibinfo {author} {\bibfnamefont {G.~I.}\ \bibnamefont
			{Martone}}\ and\ \bibinfo {author} {\bibfnamefont {S.}~\bibnamefont
			{Stringari}},\ }\bibfield  {title} {\bibinfo {title} {{Supersolid phase of a
				spin-orbit-coupled Bose-Einstein condensate: A perturbation approach}},\
	}\href {https://doi.org/10.21468/SciPostPhys.11.5.092} {\bibfield  {journal}
		{\bibinfo  {journal} {SciPost Phys.}\ }\textbf {\bibinfo {volume} {11}},\
		\bibinfo {pages} {092} (\bibinfo {year} {2021})}\BibitemShut {NoStop}%
\end{thebibliography}
\end{document}